\documentclass[aps,pra,reprint,preprintnumbers,amsmath,amssymb,floatfix]{revtex4-1}

\usepackage[utf8]{inputenc}
\usepackage[T1]{fontenc}
\usepackage{txfonts}
\usepackage{microtype}
\usepackage{eucal}
\usepackage{bm}
\usepackage{physics}
\usepackage{xr}
\usepackage{empheq}

\usepackage{color}
\usepackage{soul}

\usepackage{graphicx}
\usepackage{float}
\usepackage[draft]{pgf}

\usepackage{booktabs}

\usepackage[colorlinks,allcolors=blue]{hyperref}
\usepackage[capitalize]{cleveref}

\definecolor{Haakon}{rgb}{0,0,1}

\newcommand{\ql}{{\ve{q}\lambda}}
\newcommand{\mql}{{-\ve{q}\lambda}}

\newcommand{\ve}[1]{\boldsymbol{#1}}

\newcommand{\x}{\lambda}  

\newcommand{\xh}{\ve{\hat{x}}}
\newcommand{\yh}{\ve{\hat{y}}}
\newcommand{\zh}{\ve{\hat{z}}}

\newcommand{\nodag}{{\mathstrut}}

\newcommand*{\citen}[1]{%
  \begingroup
    \romannumeral-`\x 
    \setcitestyle{numbers,square}%
    \cite{#1}%
  \endgroup   
}

\begin{document}
\title{Magnon-polarons in cubic collinear Antiferromagnets}
\author{Haakon T. Simensen}
\author{Roberto E. Troncoso}
\author{Akashdeep Kamra}
\author{Arne Brataas}
\affiliation{Center for Quantum Spintronics, Department of Physics, Norwegian University of Science and Technology, NO-7491 Trondheim, Norway}

\date{\today}
 
\begin{abstract}
We present a theoretical study of excitations formed by hybridization between magnons and phonons - magnon-polarons -  in antiferromagnets. We first outline a general approach to determining which magnon and phonon modes can and cannot hybridize in a system thereby addressing the qualitative questions concerning magnon-polaron formation. As a specific and experimentally relevant case, we study Nickel Oxide quantitatively and find perfect agreement with the qualitative analysis, thereby highlighting the strength of the former. We find that there are two distinct features of antiferromagnetic magnon-polarons which differ from the ferromagnetic ones. First, hybridization between magnons and the longitudinal phonon modes is expected in many cubic antiferromagnetic structures. Second, we find that the very existence of certain hybridizations can be controlled via an external magnetic field, an effect which comes in addition to the ability to move the magnon modes relative to the phonons modes.
\end{abstract}

\maketitle

\section{Introduction}

Ultrafast magnetization dynamics is expected to play a pivotal role in the development of Terahertz (THz) technologies \cite{WalowskiJAP2016,MittlemanJAP2017}. These processes cover the elusive THz frequency gap that roughly spans from 100 GHz to 30 THz \cite{TH_roadmap,Tonouchi2007}, the upper and lower limits of microwave and optical techniques, respectively. An early attempt on ultrafast magnetization dynamics was made in ferromagnetic Nickel almost 20 years ago using THz lasers \cite{BeaurepairePRL1996}. Several new findings have propelled the field \cite{KirilyukRMP2010,KoopmansPRL2005}, such as ultrafast coherent control of spin waves \cite{Kampfrath2011}, ultrafast optically induced magnetization \cite{JuPRL2004,BattiatoPRL2010,BerrittaPRL2016}, magnetic switching \cite{StanciuPRL2007,Radu2011,Stupakiewicz2017} and heat-assisted magnetic recording \cite{SanderJPD2017}. The complex interplay among charge, spin, elastic and optical degrees of freedom underlies the rich physics governing the ultrafast magnetization dynamics.

Antiferromagnetic materials provide a natural niche in this field \cite{Ivanov2014,Antiferromagnets2016,Baltz2018,Jungwirth2018,Gomonay2018}. Their fast magnetization dynamics, with the potential to cover the Terahertz range \cite{Baltz2018,TH_roadmap}, and the lack of net magnetic moment \cite{Baltz2018}, have instigated a growing interest in antiferromagnets (AFMs). AFM insulators are particularly interesting due to the absence of Joule heating caused by the scattering of charge currents \cite{LinPRL2016,WuPRL2016,BenderPRL2017,LebrunNature2018}. The focus is rather on the spin currents carried by magnons, the quantized excitations of the magnetization dynamics. The ultrafast magnetization dynamics in AFM insulators can thus be understood in terms of these magnons and their interaction with the phonons - the quantized excitations of the lattice vibrations.

Although the exchange interaction underlies magnetism, the magnon-phonon interaction is crucial for the dynamics and equilibration of the spin system~\cite{Akhiezer1968}. The latter is brought on by the non-linear processes in which magnons scatter while absorbing or emitting phonons~\cite{Akhiezer1968}. In contrast, the linear magnetoelastic coupling~\cite{Kittel1949,Akhiezer1968} results in magnons and phonons combining to form hybrid excitations - magnon-polarons~\cite{Kamra2014_Actuation,Kamra2015} - when their coupled dispersions anticross. Despite the wavevector range corresponding to significant hybridization being small, it has been found to act as an energy short-circuit between the magnon and phonon subsystems~\cite{Ruckriegel2014,Bozhko2017}. While the magnon-polarons have been studied in great detail~\cite{Kittel1949,Kittel1958,Akhiezer1968}, there has been a rekindling of interest in the phenomenon due to recent advances in fabrication and measurement techniques as well as fresh breakthroughs in the field of spintronics~\cite{Chumak2015,Uchida2011,Saitoh2006}. Despite decades of study, key questions like which phonon mode should or should not hybridize with magnons remain insufficiently understood. A related question is the issue of spin conservation in magnets~\cite{Holanda2018}, which is often invoked to understand several phenomena including the formation of magnon-polarons. In particular, it has been shown
that spin conservation may not be invoked in a simple manner while addressing the effects of magnetoelastic interaction~\cite{Kamra2015}. This is because the latter is primarily rooted in spin-nonconserving interactions such as spin-orbit and dipolar contributions~\cite{Kittel1949,Akhiezer1968}. 

Magnon-polarons formed in the ferrimagnetic insulator yttrium iron garnet (YIG) have revealed their direct signatures in several recent experiments \cite{Flebus2017,Ruckriegel2014,ManPRB2017}. Pronounced features, attributed to magnon-polarons, in the magnetic field dependence of the spin Seebeck effect (SSE) \cite{Kikkawa2016,Flebus2017} as well as nonlocal spin transport \cite{Cornelissen2017} have been observed in YIG. The groundwork for these observations was laid out in previous works. For instance, coherent elastic waves were used in spin pumping \cite{Weiler2012,Uchida2011} experiments, as well as spin wave excitations \cite{Kamra2015,Li2017}. AFMs represent a step forward in this field. Although the ultrafast-response of AFM insulators is an exciting property, its control with magnetic fields is challenging. Thus its manipulation via magnetoelastic effects presents a useful alternative.

Here, we present a theoretical study of coupled magnetoelastic modes in AFMs. As compared to their ferromagnetic counterparts, the multisublattice nature of AFMs hosts qualitatively new, subtle, and rich magnetoelastic phenomena. Focusing on magnon-polaron formation, we outline a general method for gaining qualitative, physical insight into which magnon and phonon modes hybridize, given a crystal symmetry and ground state. As a special case, we study the widely used Nickel Oxide (NiO) in its collinear ground state, in which its spins are oriented at an angle with the crystal axes. This misalignment is found to permit novel effects such as hybridization with longitudinal phonons propagating along a crystal axis. Furthermore, NiO is found to host linearly polarized magnons which attain an increasingly elliptical polarization on application of an external magnetic field. This control of the qualitative nature of the magnons permits a magnetic field control of whether or not the magnons participate in forming magnon-polarons. This tunability goes well beyond the Zeeman energy shift afforded by the ferromagnetic magnons~\cite{Kikkawa2016,Flebus2017}, and opens prospects for novel functionalities. 

This paper is structured as follows. In Section \ref{sec:semiclas}, we present a general analysis of magnon-polarons in AFMs determining which phonons do or do not hybridize with which magnons. Considering NiO as an apt example, we examine its eigenmodes quantitatively finding them to be consistent with the general analysis. In Sec. \ref{sec:theory}, we derive the quantum theory describing magnons, phonons and the magnon-phonon interaction. The quantitative results of the magnon-polarons states in NiO follows in Sec. \ref{sec:results}, which in turn are compared with qualitative results predicted in Sec. \ref{sec:semiclas}. Finally, we end with discussions and conclusions in Sec. \ref{sec:discussion} and \ref{sec:conclusion}, respectively.
\section{Semi-classical qualitative analysis}
\label{sec:semiclas}

In the continuum (long-wavelength) limit, the magnetoelastic Hamiltonian in cubic AFMs is given as

\begin{align}
\begin{split}
\mathcal{H}_{\rm ME}^{\rm AFM} &= \sum_{\alpha\beta} \int \mathrm{d}^3 r B^{\alpha\beta} n^{\alpha}(\ve{r}) n^{\beta}(\ve{r}) \epsilon^{\alpha\beta}(\ve{r}) \\
&+ \sum_{\alpha\beta} \int \mathrm{d}^3 r \tilde{B}^{\alpha\beta} n^{\alpha}(\ve{r}) n^{\beta}(\ve{r}) \tilde{\epsilon}^{\alpha\beta}(\ve{r}),
\label{eq:Hafm_nontrivial}
\end{split}
\end{align}

\noindent where $\epsilon^{\alpha\beta} = \frac{1}{2} \left( \frac{\partial u^\alpha}{\partial r^\beta} + \frac{\partial u^\beta}{\partial r^\alpha} \right)$ is the strain tensor, where $u^\alpha$ is the atom displacement field in the $\alpha$ direction. ${B^{\alpha\beta} = B^\parallel \delta^{\alpha\beta} + B^\perp \left(1 - \delta^{\alpha\beta} \right)}$ and ${\tilde{B}^{\alpha\beta} = \tilde{B}^\parallel \delta^{\alpha\beta} + \tilde{B}^\perp \left(1 - \delta^{\alpha\beta} \right)}$ are 4 magnetoelastic coefficients, and $\ve{n}$ is the Néel field. $\tilde{\epsilon}$ is an elastic tensor with elements being linear combinations of elements of the strain tensor. The first term is the antiferromagnetic analogy of the conventional ferromagnetic magnetoelastic Hamiltonian \cite{Kittel1949}. The second term, where $\tilde{\epsilon}$ appears, derives from the internal spin structure. This term is unique to materials with at least two sublattices, and the exact form of $\tilde{\epsilon}$ depends upon the spin structure of the AFM. If the spin structure is trivial, meaning that all $n$'th nearest neighbors of a lattice site belong to a single sublattice for any $n$, then this term disappears. We will refer to spin structures which do not fulfill this requirement as being non-trivial. A full derivation of this Hamiltonian is given in Appendix \ref{sec:app3}.

To start with we will consider a simple AFM with a trivial spin structure, where only the first term in the antiferromagnetic magnetoelastic Hamiltonian \eqref{eq:Hafm_nontrivial} appears. We will only consider magnons and phonons propagating along one of the crystal axes, which we define to be the $\zh$ direction for concreteness. The three independent phonon modes are then proportional to $\partial u^{\gamma}/\partial r^{z}$, where $\gamma \in \{ x,y \}$ describe transverse phonons, whereas $\gamma = z$ describes a longitudinal phonon.

In order to give a physical interpretation of the hybridization, we need to express the Hamiltonian in terms of both the free magnon and phonon eigenmodes. The strain tensor components $\epsilon^{\alpha\beta}$ are superpositions of phonon eigenmodes, and the Hamiltonian is thus already given in terms of phonon eigenmodes. The remaining task is therefore to find the magnon eigenmodes expressed as function of the Néel field.

\subsubsection{Circularly polarized magnons in easy axis antiferromagnets}
Let us now consider an easy axis AFM where the spins align along the $\zh$ axis in the (classical) ground state. In easy-axis AFMs the Hamiltonian is invariant to a global spin rotation about the $\zh$ axis, and we therefore expect the magnons modes to be circularly polarized. Moreover, as magnons corresponds to small deviations from the spin ground state, magnons leave $n^z$ approximately constant, whereas $n^x$ and $n^y$ are expected to oscillate harmonically. These considerations combined imply that we may express the magnon modes as $\alpha \equiv n_x + i n_y$ and $\beta \equiv n_x - i n_y$. The magnetoelastic Hamiltonian can then be rewritten as

\begin{equation}
\mathcal{H}_{\rm ME}^{\rm circ} = \frac{B^\perp n^z}{4}  \left[ \alpha \left( \frac{\partial u^x}{\partial r^z} - i \frac{\partial u^y}{\partial r^z} \right) +  \beta \left( \frac{\partial u^x}{\partial r^z} + i \frac{\partial u^y}{\partial r^z} \right)  \right].
\label{eq:afm_easyaxis}
\end{equation}

\noindent Note that in this situation, angular momentum in the $\zh$ direction is conserved. If we create a magnon, this may hybridize and produce a circularly polarized phonon, which has got angular momentum along $\zh$. If we create a linearly polarized phonon (with zero angular momentum along $\zh$), it may hybridize and produce spinless combination of magnons $\propto \alpha \pm \beta$. Angular momentum is hence conserved. This follows directly from the rotational symmetry about the $\zh$ axis \footnote{Angular momentum conservation requires a continuous rotational symmetry about the $\zh$ axis. We consider cubic symmetry, that is a finite symmetry group, which in principle is not sufficient for such an angular momentum conservation argument. However, to second order in phonons and magnons, this argument holds anyway. This is because we neglect the effect of $(n^x)^2$, $(n^y)^2$ and $(n^z)^2$, which would produce third order terms in magnons and phonons. That is, $B^\parallel$ is not included in the equations, and we are in principle free to set this to any value, as it does not affect the physics. Set it to $B^\parallel = B^\perp$, and we have in fact assumed isotropic symmetry.}.

\subsubsection{Linearly polarized magnons in biaxial NiO}
\label{sssec:biaxialNiO}
We will now consider an AFM with two hard-axis anisotropies. For concreteness we will use NiO as an example. To start with, we will neglect its non-trivial spin structure, which introduces the second term in the magnetoelastic Hamiltonian \eqref{eq:Hafm_nontrivial}. We will therefore solely focus on the first term in the Hamiltonian \eqref{eq:Hafm_nontrivial}, where only the conventional strain tensor appears. Following the derivation of the Hamiltonian in Appendix \ref{sec:app3}, we find that this is equivalent to assuming that next nearest neighbor interaction is the dominant term contributing to the magnetoelastic interaction.

We start once again from the Hamiltonian \eqref{eq:Hafm_nontrivial}, which we want to express in terms of the magnon eigenmodes. First, we need to rotate the coordinate system so that the new $\zh'$ axis coincides with the spin condensation axis. The spins in NiO condense along one of the 12 equivalent $[\bar{1}\bar{1}2]$ directions, within internally ferromagnetic [111] planes. We therefore define the primed coordinate system as $\zh' = \frac{1}{\sqrt{6}}[-1,\, -1,\, 2]$, $\xh' = \frac{1}{\sqrt{3}} [1 ,\, 1,\, 1]$ and $\yh' = \frac{1}{\sqrt{2}} [-1 ,\, 1,\, 0]$. Define the rotation matrix $\mathbf{O}$ so that the primed and un-primed coordinate systems are related by $\ve{r} = \mathbf{O} \ve{r}'$. By writing $\ve{n} = \mathbf{O} \ve{n}'$, we find

\begin{align}
\begin{split}
\mathcal{H}_{\rm ME}^{\rm NiO} &= \frac{n'^x n'^z}{3 \sqrt{2}} \left[ 2 B^\parallel  \frac{\partial u^z}{\partial r^z} + \frac{B^\perp}{2} \left(  \frac{\partial u^x}{\partial r^z} + \frac{\partial u^y}{\partial r^z}  \right) \right] \\
&+  \frac{n'^y n'^z}{2 \sqrt{3}}  B^\perp \left(\frac{\partial u^y}{\partial r^z} - \frac{\partial u^x}{\partial r^z} \right),
\label{eq:afm_nio}
\end{split}
\end{align}

\noindent where we once again have assumed that the magnon and phonons propagate along the $\zh$ direction.

In NiO, the magnon eigenmodes are linearly polarized and spinless. In other words, the magnon eigenmodes correspond semi-clasically to oscillations of $n'^x$ and $n'^y$ separately. Hence, the Hamiltonian \eqref{eq:afm_nio} is in fact already given in terms of both the magnon and phonon eigenmodes, and can thus be directly interpreted. We consider first the transverse phonons. Both magnon modes hybridize with both transverse phonon modes, however with different interaction parameters. As a consequence of this, angular momentum in the $\zh$ direction is no longer conserved. This is a direct consequence of the lack of rotational symmetry about the $\zh$ axis due to the spin condensation axis $\zh'$ not being aligned with the momentum direction $\zh$. Further, we note that only the $n'^x$ magnon mode hybridizes with the longitudinal phonon mode $\partial u^z/\partial r^z$. The $n'^x$ mode is the mode oscillating along the axis with the largest anisotropy, and is followingly the most energetic mode. We therefore conclude that the lower magnon mode passes the longitudinal phonon modes undisturbed, while the upper one is expected to hybridize.

Note that the exact decoupling of the lower magnon mode from the longitudinal phonon modes in the Hamiltonian \eqref{eq:afm_nio} is a consequence of the magnon eigenmodes being linearly polarized. If the magnon eigenmodes were not linearly polarized, the longitudinal phonon mode would in general couple to both phonon modes. This can be realized by applying an external magnetic field along the $\zh'$ axis. The effect is that the magnon eigenmodes can be described semi-classically as elliptical precessions of the Néel field around the ground state. For concreteness, let us assume that the eigenmodes are elliptically polarized, $\alpha = \left( A n'^x + i B n'^y \right)$ and $\beta = \left( B n'^x - i A n'^y \right)$ \footnote{The qualitative results do not depend on the exact form of the eigenmodes as long as they mix $n'^x$ and $n'^y$.}, where $A$ and $B$ depend on the magnetic field strength. The Hamiltonian \eqref{eq:afm_nio} expressed in terms of the eigenmodes then follows as

\begin{align}
\begin{split}
\mathcal{H}_{\rm ME}^{\rm NiO} &= \left\{ \left( A \alpha + B \beta \right) \left( \frac{1}{3 \sqrt{2}} \right) \left[ 2 B^\parallel  \frac{\partial u^z}{\partial r^z} + \frac{B^\perp}{2} \left(  \frac{\partial u^x}{\partial r^z} + \frac{\partial u^y}{\partial r^z}  \right) \right]  \right. \\
&+ \left( A \beta - B \alpha  \right) \left[ \left. \frac{i}{2 \sqrt{3}}  B^\perp \left(\frac{\partial u^y}{\partial r^z} - \frac{\partial u^x}{\partial r^z} \right) \right]
 \right\} \times \frac{n'^z}{(A^2 + B^2)}.
\label{eq:afm_ellipse}
\end{split}
\end{align}

\noindent As expected, both magnons $\alpha$ and $\beta$ now hybridize with the longitudinal phonon $\partial u^z/\partial r^z$.

\subsubsection{Antiferromagnets with internally ferromagnetic planes}
\label{sssec:biaxialNiO_nontrivial}
In the last section, we considered magnon-phonon hybridization in NiO under the assumption that the second term in the Hamiltonian \eqref{eq:Hafm_nontrivial} is negligible. We will now look at the effect of the second, spin structure dependent term. If we include only those terms contributing to the magnon-phonon hybridization with momentum along the $\zh$ direction \footnote{Additional strain tensor elements appear in the spin structure dependent tensor if we do not require momentum conservation. Additional elements also appear if we consider uniform magnetostriction.}, the spin structure dependent tensor in NiO is

\begin{align}
\begin{split}
\tilde{\bm{\epsilon}} =
\begin{pmatrix}
\epsilon^{xz} & \epsilon^{xz} + \epsilon^{yz} & \epsilon^{zz} \\
\epsilon^{xz} + \epsilon^{yz} & \epsilon^{yz} & \epsilon^{zz} \\
\epsilon^{zz} & \epsilon^{zz} & 0 \\
\end{pmatrix}.
\end{split}
\label{eq:eps_tilde}
\end{align}

\noindent The further process of interpreting the hybridization is just the same as shown above; rewrite the Hamiltonian in terms of the magnon eigenmodes, and then read off which modes hybridize. The result is

\begin{align}
\begin{split}
\mathcal{H}_{\rm ME}^{\rm NiO} &= \frac{\sqrt{2} n'^x n'^z}{3} \left[ 2 \tilde{B}^\perp  \frac{\partial u^z}{\partial r^z} - \frac{(\tilde{B}^\parallel + \tilde{B}^\perp)}{2} \left(  \frac{\partial u^x}{\partial r^z} + \frac{\partial u^y}{\partial r^z}  \right) \right] \\
&+  \frac{n'^y n'^z}{\sqrt{3}}  \tilde{B}^\parallel \left(\frac{\partial u^x}{\partial r^z} - \frac{\partial u^y}{\partial r^z} \right).
\label{eq:afm_nio_nontrivial}
\end{split}
\end{align}

\noindent Evidently, the spin structure dependent term in the Hamiltonian does not introduce any new types of hybridization, as the magnon and phonon modes which couple are identical to those appearing in Eq. \eqref{eq:afm_nio}. Therefore, the discussion of a simplified NiO-like material in Sec. \ref{sssec:biaxialNiO} appears to be valid for the real NiO as well. We predict that the decoupling between a magnon mode and the longitudinal modes can be lifted by applying an external magnetic field along the $\zh'$ axis. The magnetic field can also be used to smoothly tune the hybridization between the modes, as the coefficients $A$ and $B$ appearing in Eq. \eqref{eq:afm_ellipse} depend on the field strength.

\section{Quantized Hamiltonian}
\label{sec:theory}
In this section, we will derive the quantized Hamiltonian which is later used to find the exact magnon-polarons in NiO. We start by deriving the magnon Hamiltonian, followed by the phonon Hamiltonian. Last, we will derive the terms which couple magnons and phonons into an effective hybridized state. We stress that although the exact derivations which follow are specific to NiO, the method is fully general and hence valid for all collinear AFMs.

Above the Néel temperature, NiO forms the FCC-structure, whereas it is slightly distorted into a rhombohedral one below \cite{Rooksby1948}. This distortion from cubic symmetry is very small, corresponding to an angle of about $0.07^\circ$ \cite{Slack1960}, and we will therefore neglect it in the following derivations.

\begin{figure}[htb!]
\centering
\includegraphics[width=0.9\columnwidth]{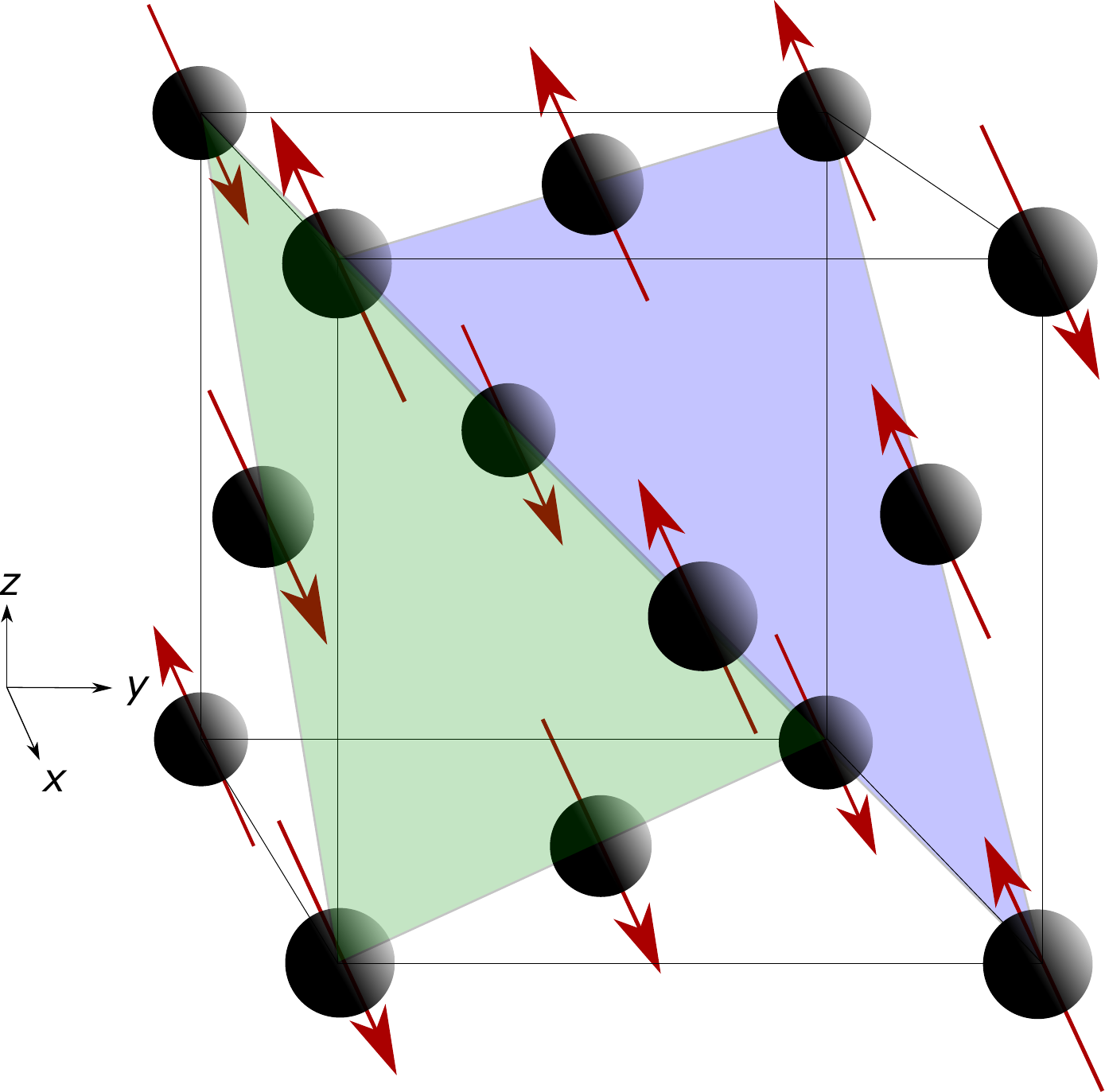}
\caption{Spin configuration in one of the (classical) ground states of NiO. The ferromagnetic planes are in the $[111]$ direction with spins along the $[\bar{1}\bar{1}2]$ direction. Only the magnetic $\mathrm{Ni}^{2+}$ ions are depicted. The green and blue planes mark the two different sublattices.}
\label{fig:NiO}
\end{figure}

\subsection{Magnons}
\label{ssec:magnons}
The antiferromagnetic ordering in NiO is well established. Below the Néel temperature, the spins order in internally ferromagnetic $(111)$ planes \cite{Shull1951, Hutchings1972}. The spins on two such neighboring planes are antiparallel, causing the overall structure to be antiferromagnetic. Due to the cubic symmetry of the FCC structure, there are four equivalent $(111)$ planes. In each plane there is moreover a 3-fold degeneracy in the spin direction. There are thus 12 possible antiferromagnetic ground states, one of which are depicted in Fig. \ref{fig:NiO}. As magnons are small spin fluctuations relative to a ground state, we may choose to work from any one of these twelve possible ground states without loss of generalization. We here choose the $(111)$ plane as the plane of internally ferromagnetic order, and $\zh' = \frac{1}{\sqrt{6}}[-1,\, -1,\, 2]$ as the spin quantization axis along which the spins in the (classical) antiferromagnetic ground state are aligned. We moreover let $\xh' = \frac{1}{\sqrt{3}} [1 ,\, 1,\, 1]$ and $\yh' = \frac{1}{\sqrt{2}} [-1 ,\, 1,\, 0]$ define the rest of the primed coordinate system. Note that $\xh'$ is perpendicular to the $(111)$ plane, and $\yh'$ is parallel to it. The spins form a bipartite lattice, where the sublattice with spin up (down) is named sublattice A (B). 

In the magnetic Hamiltonian we will include exchange interaction and two hard-axis anisotropies. In order to obtain a sufficiently accurate dispersion relation for magnons in NiO, we need to include exchange coupling between both nearest neighbour (nn) and next-nearest neighbour (nnn) spins. Every spin site has 6 nn's on the same sublattice and 6 nn's on the opposite sublattice, as well as 6 nnn's on the opposite sublattice. The magnetic Hamiltonian follows as \cite{Lines1965,Hutchings1972}

\begin{align}
\begin{split}
\mathcal{H}_{\rm m} &= 
\frac{1}{2} \sum_{i, \ve{\delta}_1} J_{1} \ve{S}_i \cdotp \ve{S}_{i + \ve{\delta}_1}
+
\frac{1}{2} \sum_{i, \ve{\delta}_2} J_{2} \ve{S}_i \cdotp \ve{S}_{i+ \ve{\delta}_2} \\
&+
\sum_{i} D_{x'} \left( S_i^{x'} \right)^{2}
+
\sum_{i} D_{y'} \left( S_i^{y'} \right)^{2},
\end{split}
\end{align}

\noindent where $\ve{S}_i^{\alpha'}$ refer to the spin component in the $\alpha'$ direction at lattice site $i$, and $D_{x'} > 0$ and $D_{y'} > 0$ are anisotropy constants. The summation index $i$ runs over the whole lattice, and $\ve{\delta}_1$ and $\ve{\delta}_2$ run over nn's and nnn's to lattice site $i$, respectively. $J_1$ and $J_2$ are the corresponding exchange coupling constants. We split the $i$ summations into sums over sublattices A and B with spins $\ve{S}^\mathrm{A}$ and $\ve{S}^\mathrm{B}$, and do a Holstein-Primakoff transformation of the spin operators in terms of boson operators $a$ and $b$:

\begin{align}
\begin{split}
S^{\mathrm{A} z'}_{i} &= S - a_{i}^{\dagger} a_{i}^{\nodag}, \\
S^{\mathrm{A} +}_{i} &= \sqrt{2S} a_{i}^{\nodag}, \\
S^{\mathrm{A} -}_{i} &= \sqrt{2S} a_{i}^{\dagger},   \\
\end{split}
\begin{split}
S^{\mathrm{B} z'}_{j} &= - S + b_{j}^{\dagger} b_{j}^{\nodag}, \\
S^{\mathrm{B} +}_{j} &= \sqrt{2S} b_{j}^{\dagger}, \\
S^{\mathrm{B} -}_{j} &= \sqrt{2S} b_{j}^{\nodag},
\end{split}
\end{align}

\noindent where we have assumed that ${\langle a_{i}^{\dagger} a_{i}^{\nodag}\rangle/ 2S \ll 1}$ and ${\langle b_{j}^{\dagger} b_{j}^{\nodag}\rangle/ 2S \ll 1}$, and $S = 1$ in NiO. We then perform a Fourier transformation of the operators

\begin{align}
\begin{split}
a_i^{\nodag} &= \frac{1}{\sqrt{N_A}} \sum_{\ve{k}} e^{-i \ve{k} \cdotp \ve{x_i}} a_{\ve{k}}^{\nodag}, \\
a_i^{\dagger} &= \frac{1}{\sqrt{N_A}} \sum_{\ve{k}} e^{i \ve{k} \cdotp \ve{x_i}} a_{\ve{k}}^{\dagger}, \\
\end{split}
\begin{split}
b_j^{\nodag} &= \frac{1}{\sqrt{N_B}} \sum_{\ve{k}} e^{-i \ve{k} \cdotp \ve{x_j}} b_{\ve{k}}^{\nodag}, \\
b_j^{\dagger} &= \frac{1}{\sqrt{N_B}} \sum_{\ve{k}} e^{i \ve{k} \cdotp \ve{x_j}} b_{\ve{k}}^{\dagger}, \\
\end{split}
\end{align}

\noindent where $\ve{x}_i$ and $\ve{x}_j$ are position vectors on sublattice A and B. Now let $\sum_{\ve{\delta}_{n} \in ab}$ denote the sum over the $n$'th nearest neighbors on sublattice $b$ of a spin belonging to sublattice $a$, where $a,b \in \{\mathrm{A},\,\mathrm{B} \}$. Let $z_{n}^{ab}$ be the number of such neighbors. Use this definition to define the quantity

\begin{equation}
\gamma^{ab}_{n\ve{k}} = \sum_{\ve{\delta}_{n} \in ab} e^{i \ve{k} \cdotp \ve{\delta}_{n}}.
\end{equation}

\noindent One can then show that the Hamiltonian takes the form

\begin{align}
\begin{split}
\mathcal{H}_{\rm m} = &\sum_{\ve{k}} \Bigg[ A_{\ve{k}} \left( a_{\ve{k}}^{\dagger} a_{\ve{k}}^{\nodag} + b_{\ve{k}}^{\dagger} b_{\ve{k}}^{\nodag} \right) + B_{\ve{k}} \left( a_{\ve{k}}^{\nodag} b_{-\ve{k}}^{\nodag} + a_{\ve{k}}^{\dagger} b_{-\ve{k}}^{\dagger}  \right) \\
&\quad + C \left( a_{\ve{k}}^{\nodag} a_{-\ve{k}}^{\nodag} + b_{\ve{k}}^{\nodag} b_{-\ve{k}}^{\nodag}   \right) + C \left( a_{\ve{k}}^{\dagger} a_{-\ve{k}}^{\dagger} + b_{\ve{k}}^{\dagger} b_{-\ve{k}}^{\dagger}  \right) \Bigg],
\end{split}
\label{eq:H_magnon}
\end{align}

\noindent where we have introduced the following coefficients

\begin{align}
A_{\ve{k}} &= J_1 S\gamma^{\rm AA}_{1\ve{k}} + J_2 S z_{2}^{AB} + S \left( D_{x'} + D_{y'} \right), \\
B_{\ve{k}} &= J_1 S \gamma^{AB}_{1\ve{k}} + J_2 S \gamma^{AB}_{2\ve{k}}, \\
C &= \frac{S}{2} \left( D_{x'} - D_{y'} \right).
\end{align}

\noindent This boson Hamiltonian can be diagonalized following the procedure of Ref. \citen{White1965}. We define

\begin{align}
\begin{split}
\ve{\xi}_{\ve{k}}^{\nodag} =
\begin{pmatrix}
a_{\ve{k}}^{\nodag} \\
b_{-\ve{k}}^{\dagger} \\
a_{-\ve{k}}^{\dagger} \\
b_{\ve{k}}^{\nodag} \\
\end{pmatrix},
\end{split}
\begin{split}
\mathbf{H}_{\ve{k}}^{\rm m} = 
\begin{pmatrix}
A_{\ve{k}} & B_{\ve{k}} & 2 C & 0 \\
B_{\ve{k}} & A_{\ve{k}} & 0 & 2 C \\
2 C & 0 & A_{\ve{k}} & B_{\ve{k}} \\
0 & 2 C & B_{\ve{k}} & A_{\ve{k}}
\end{pmatrix},
\end{split}
\end{align}

\noindent so that the Hamiltonian takes the form

\begin{equation}
\mathcal{H}_{\rm m} = \frac{1}{2} \sum_{\ve{k}} \xi_{\ve{k}}^{\dagger} \mathbf{H}_{\ve{k}}^{\rm m} \xi_{\ve{k}}^{\nodag}.
\label{eq:Hmag}
\end{equation}

\noindent Now define $\phi_{\ve{k}} = \mathbf{T}_{\ve{k}}^{-1} \xi_{\ve{k}}$ as the vector of operators that by definition diagonalizes the Hamiltonian. Note that $\mathbf{T}_{\ve{k}}$ in general is a non-unitary transformation, that is $\mathbf{T}_{\ve{k}}^\dagger \neq \mathbf{T}_{\ve{k}}^{-1}$. Now define a matrix $\mathbf{g}$ as a commutator between the vector of (bosonic) operators and its Hermitian adjoint, $\mathbf{g} \equiv \comm{ \xi_{\ve{k}}^{\nodag}}{ \xi_{\ve{k}}^{\dagger}}$. By inserting $\xi_{\ve{k}} = \mathbf{T} \phi_{\ve{k}}$ into the commutator, one can show that the tranformation matrix $\mathbf{T}$ satisfies $\mathbf{T}^\dagger = \mathbf{g}^{-1} \mathbf{T}^{-1} \mathbf{g}$. The Hamiltonian can then be written into the eigenvalue equation

\begin{equation}
\mathbf{g} \mathbf{H}_{\ve{k}}^{\rm m} \mathbf{T}_{\ve{k} i} = \hbar \omega_{\ve{k}} g_{ii} \mathbf{T}_{\ve{k} i},
\end{equation}

\noindent where $\mathbf{T}_{\ve{k} i}$ is the $i$'th row of $\mathbf{T}_{\ve{k}}$, and $\hbar \omega_{\ve{k}}$ is the energy of the magnon mode $\ve{k}$. Solving this eigenvalue equation results in 

\begin{equation}
\hbar \omega_{\ve{k}}^{\pm} = \frac{1}{2} \sqrt{A_{\ve{k}}^2 - \left( B_{\ve{k}}^{\nodag} \pm 2 C \right)^2},
\end{equation}

\noindent which describes the dispersion relations for the two free magnon modes in NiO. Experimentally fitted values for the parameters can be looked up in for instance Ref. \citen{Hutchings1972}.

\subsection{Phonons}
NiO forms the FCC structure with two atoms in the basis, one nickel atom and one oxygen atom. These are separated by a distance $a/2$, where $a \approx 4.17 \; \AA$ \cite{Bartel1971} is the cubic lattice constant. Let $s$ be an index referring to the atom type, with Ni as type 1 and O as type 2. Moreover, let $i$ refer to a site on the FCC lattice at which one such pair of atoms belong. The indices $(i,s)$ therefore uniquely refer to one specific atom in the structure. Now let $u_{is}^{\alpha}$ be the displacement from equilibrium of atom $(i,s)$ in the $\alpha$ direction, where $\alpha \in \{x, y, z \}$. Finally, let $K_{is,jt}^{\alpha\beta}$ be the proportionality constant between the force acting on atom $(i,s)$ in the $\alpha$-direction and the displacement of atom $(j,t)$ in the $\beta$-direction. The phonon energies are then given as the solution of the classical equation of motion

\begin{equation}
-\omega^2 m_{s} u_{is}^{\alpha} = \sum_{j} \sum_{t} \sum_{\beta} K_{is,jt}^{\alpha\beta} u_{jt}^{\beta}.
\label{eq:eqofmot}
\end{equation}

\noindent This equation has $3 n$ momentum dependent solutions: $3$ acoustic and $3(n - 1)$ optical modes, where $n$ is the number of atoms in the basis. With $n=2$ in NiO, this results in a total of 6 phonon modes. We name these different modes $\omega_{\ve{q}\lambda}$, where $\ve{q}$ is the phonon momentum and $\lambda$ labels the mode. The quantized phonon Hamiltonian follows as

\begin{equation}
\mathcal{H}_{\rm p} = \sum_{\ve{q}\lambda} \hbar \omega_{\ve{q}\lambda} c_{\ve{q}\lambda}^{\dagger}c_{\ve{q}\lambda}^{\nodag},
\label{eq:Hph}
\end{equation}

\noindent where $c_{\ve{q}\lambda}^{\nodag}$ and $c_{\ve{q}\lambda}^{\dagger}$ are phonon annihilation and creation operators.

Eq. \eqref{eq:eqofmot} is a very general equation, where the phonon energies can be found straightforwardly given a set of force coefficients $K_{is,jt}^{\alpha\beta}$. We will here use the rigid-ion model to compute the phonon eigenmodes in NiO. The rigid-ion model \cite{Kellermann1940,Cochran1971} is perhaps the simplest model which is able to reproduce a relatively accurate picture of phonons in NiO. It is not the most accurate available model, but it has the advantage of having only a few adjustable parameters, and will suffice for our usage. It is based on modelling each atom as a rigid sphere which moves around an equilibrium position, and is well explained in Refs. \citen{Kellermann1940,Cochran1971}. Each atom is connected to its nearby atoms by springs which represent the short-ranged forces between nearby atoms. We include energy terms linear and quadratic in the Ni-O displacements (with force constants $B_{12}$ and $A_{12}$ respectively) and O-O displacements (with force constants $B_{22}$ and $A_{22}$ respectively), while we may neglect the explicit Ni-Ni displacement terms \cite{Reichardt1975}. 

Additionally, we need to include Coulomb interactions in order to get a realistic model of the optical phonons \footnote{The energy difference between the transverse and longitudinal optical modes depends solely on the effective charge $Z \abs{e}$. As this energy difference is about 5 THz (in units of $2\pi \hbar$) in NiO, we obviously need to include Coulomb forces in order to attain a realistic phonon dispersion.}. Each atom is given an effective charge $\pm Z \abs{e}$. The Coulomb interaction is long-ranged, meaning that interactions between atoms infinitely far apart contribute. An infinite sum obviously causes numerical difficulties, and to solve this we use a so-called Ewald summation; we split the real space Coulomb summation into a real space sum and a Fourier space integral. We sum over the closest atoms in real space, and approximate the sum over more distant atoms by an integral in Fourier space. This enables us to approximate the formally infinite sum by summing over about $10$ lattice sites in real space. We eventually fit the five constants $A_{12}$, $A_{22}$, $B_{12}$, $B_{22}$ and $Z$ to an experimentally measured phonon dispersion in Ref. \citen{Reichardt1975}. The formal details of this calculation apart from those given above will not be covered here, as it is rather tedious, and we refer the reader to Ref. \citen{Kellermann1940} for further reading.

Each phonon mode is characterized by the atoms moving in a unique pattern. The polarization vector $\ve{\epsilon}_{\ql}$ is formally a six-component vector describing the axes along which the two atom types move for the different phonon modes $\lambda$, as well as the relative phase between the atoms types. As there are three acoustic modes and three optical ones, the only difference between the first and latter three modes is a relative phase $\pi$ between the Ni and O atoms in the optical mode. In the next section, we will use the polarization vector to couple atom displacement and spins. Since only the Ni atoms make a significant contribution to the magnetic Hamiltonian, we will only couple Ni sites. In the following, we may therefore define a three-component polarization vector $\ve{\epsilon}_{\ql}$ describing the movement of the Ni atoms only. As the polarizations of Ni atoms are identical in the acoustic and optical modes, we need only define three different polarizations.

The polarization vector must satisfy the orthogonality relation $\ve{\epsilon}_{\ve{q} \lambda}^{*} \cdotp \ve{\epsilon}_{\ve{q} \lambda'}^{\nodag} = \delta_{\lambda\lambda'}$, as well as the completeness relation $\sum_{\ve{q}} \ve{\epsilon}_{\ve{q} \lambda}^{\nodag} \ve{\epsilon}_{\ve{q} \lambda'}^{\dagger} = \mathbb{I}$. Last but not the least, the phonon polarization vectors must be eigenvectors of the equation of motion, Eq. \eqref{eq:eqofmot}. We may conveniently choose the polarization vectors such that $\ve{\epsilon}_{\ve{q}  \lambda}^{*} = \ve{\epsilon}_{-\ve{q} \lambda}^{\nodag}$ \cite{Ruckriegel2014}. We choose the polarization vectors to be \cite{Flebus2017}

\begin{align}
\begin{split}
\ve{\epsilon}_{\ve{q} 1} &= [\cos\theta_{\ve{q}}\cos\phi_{\ve{q}} ,\, \cos\theta_{\ve{q}}\sin\phi_{\ve{q}} ,\, -\sin\theta_{\ve{q}}],\\
\ve{\epsilon}_{\ve{q} 2} &= i [-\sin\phi_{\ve{q}} ,\, \cos\phi_{\ve{q}} ,\, 0],\\
\ve{\epsilon}_{\ve{q} 3} &= i [\sin\theta_{\ve{q}}\cos\phi_{\ve{q}} ,\, \sin\theta_{\ve{q}}\sin\phi_{\ve{q}} ,\, \cos\theta_{\ve{q}}],
\end{split}
\end{align}

\noindent where $\phi_{\ve{q}}$ and $\theta_{\ve{q}}$ are standard spherical coordinates defining the direction of the momentum $\ve{q}$. $\lambda = 1$ and $\lambda = 2$ describe transversal modes, while $\lambda = 3$ describes longitudinal modes. Recall that these are polarization vectors both for the acoustic and optical modes.

\subsection{Magnon-phonon coupling}
We will consider magnetoelastic coupling which hybridizes the magnon and phonon modes. Finding the new hybridized eigenstates requires us to diagonalize the Hamiltonian containing magnetic and elastic degrees of freedom. The Hamiltonian under consideration must therefore be quadratic in magnon and phonon operators, meaning that we are only to include interaction terms containing one operator of each sort. The displacement of an ion from equilibrium $\ve{u}_i$ is a measure for the elastic degree of freedom, while the spin $\ve{S}_i$ at site $i$ is a measure for the magnetic degree of freedom. $\ve{u}_i$ is linear in phonon operators (see Eq. \eqref{eq:disp}), and we may therefore immediately conclude that the interaction term must be linear in $\ve{u}_i$.

In Appendix \ref{sec:app2} we do a phenomenological expansion in spins $\ve{S}_i$ and displacements $\ve{u}_j$ to arrive at two magnetoelastic Hamiltonians. Both terms are linear in displacements $\ve{u}_j$, while they are of first and second order in spin. We show that these terms have their origin, among other things, in spin-orbit coupling between a spin and its neighboring ions' orbital momenta, and in distance dependent exchange interaction, respectively. We will henceforth assume that the term arising from the exchange interaction dominates, and we therefore neglect all terms which are not second order in spin.

The magnetoelastic Hamiltonian under consideration is therefore \cite{Evenson1969}

\begin{equation}
\mathcal{H}_{\rm ME} = \sum_{\alpha\beta\gamma\lambda} \sum_{i,\ve{\delta}} B_{i,i+\ve{\delta}}^{\alpha\beta\gamma\lambda} S_{i}^{\alpha} S_{i+\ve{\delta}}^{\beta} R_{i,i+\ve{\delta}}^{\gamma\delta},
\label{eq:HME1}
\end{equation}

\noindent where $i$ is summed over all magnetic lattice sites, $\ve{\delta}$ is a vector pointing from lattice site $i$ to one of its neighboring magnetic atoms, and $\alpha,\beta,\gamma,\lambda \in \{x,y,z\}$ refer to spatial directions. $B_{ij}^{\alpha\beta\gamma\lambda}$ is a tensor of coupling coefficients. $R_{ij}^{\gamma\lambda}$ describes local strains, and we name it the discrete strain tensor. It is defined as

\begin{equation}
R^{\gamma\lambda}_{ij} = \frac{1}{2} \frac{1}{| \ve{r}_i - \ve{r}_j |^2} \Bigg[ \left( r_i^{\gamma} - r_j^{\gamma} \right)\left( u_i^{\lambda} - u_j^{\lambda} \right) + \left( r_i^{\lambda} - r_j^{\lambda} \right)\left( u_i^{\gamma} - u_j^{\gamma} \right) \Bigg].
\label{eq:disc_strain}
\end{equation}

\noindent The discrete strain tensor simplifies to a constant times the continuous strain tensor $\epsilon^{\gamma\lambda}=\frac{1}{2} \left( \frac{\partial u^{\gamma}}{\partial r^{\lambda}} + \frac{\partial u^{\lambda}}{\partial r^{\gamma}} \right)$ in the long-wavelength limit. Note that $R_{ij}^{\alpha\beta}$ is symmetric under exchange of spatial coordinates. We could in principle have coupled the spins to an anti-symmetric elastic tensor as well. An anti-symmetric elastic tensor analogous to $R_{ij}^{\alpha\beta}$ describes local rotations. We will however disregard rotations in our analysis, and couple therefore the spins exclusively to $R_{ij}^{\gamma\lambda}$.

The number of coefficients $B_{ij}^{\alpha\beta\gamma\lambda}$ appearing in the Hamiltonian \eqref{eq:HME1} can be reduced considerably by applying Neumann's principle, stating that the Hamiltonian must be invariant under symmetry operations of the material itself \cite{Birss1964,KATZIR2004}. NiO forms the FCC structure above its Néel temperature, and its structure therefore belongs to the cubic symmetry group $O_h$. Neumann's principle states that the Hamiltonian must be invariant under symmetry operations $\mathcal{R} \in O_h$, i.e. ${ \mathcal{R}^{-1} \mathcal{H}_{\rm ME} \mathcal{R} = \mathcal{H}_{\rm ME} }$. Furthermore, the Hamiltonian must be translationally invariant. By requiring these symmetries to be fulfilled, we find that the Hamiltonian reduces to

\begin{equation}
\mathcal{H}_{\rm ME} = \sum_{\alpha\beta} \sum_{i,\ve{\delta}} B_{|\ve{\delta}|}^{\alpha\beta} S_{i}^{\alpha} S_{i+\ve{\delta}}^{\beta} R_{i,i+\ve{\delta}}^{\alpha\beta},
\label{eq:HME2}
\end{equation}

\noindent where $B_{|\ve{\delta}|}^{\alpha\beta} = \delta^{\alpha\beta} B_{|\ve{\delta}|}^{\parallel} + (1-\delta^{\alpha\beta}) B_{|\ve{\delta}|}^{\perp}$ \cite{Flebus2017}, and $\delta^{\alpha\beta}$ is the Kronecker delta. The $|\ve{\delta}|$ index of $B_{|\ve{\delta}|}^{\alpha\beta}$ means that the coefficients coupling any atoms separated by an equilibrium distance $|\ve{\delta}|$ are equal, which is due to translational and rotational invariance. In other words, there are two coefficients appearing in the Hamiltonian for every $n$'th nearest neighbor spins included in the summation over $\ve{\delta}$.

The displacement vector of a nickel atom at lattice site $i$ can be expressed in terms of the phonon operators as

\begin{equation}
\ve{u}_{i} =  \sum_{\ve{q},\lambda} \ve{\epsilon}_{\ve{q} \lambda} \sqrt{\frac{\hbar}{2 m \omega_{\ve{q}\lambda} N}} \left( c_{\ve{q},\lambda}^{\dagger} + c_{-\ve{q},\lambda} \right) e^{i \ve{q} \cdotp \ve{r}_i},
\label{eq:disp}
\end{equation}

\noindent where $\omega_{\ve{q}\lambda}$ is the angular frequency of the phonon mode $\lambda$, $m$ is the mass of the nickel atom, and $N$ is the number of nickel lattice sites. The strain tensor between two nickel atoms at position $i$ and $i + \ve{\delta}$ therefore follows as

\begin{align}
\begin{split}
R_{i,i+\ve{\delta}}^{\alpha\beta} &=  \sum_{\ve{q},\lambda} \sqrt{\frac{\hbar}{2 m \omega_{\ve{q}\lambda} N}} \left( c_{\ve{q},\lambda}^{\dagger} + c_{-\ve{q},\lambda} \right) e^{i \ve{q} \cdotp \ve{r}_i} \\
&\times \left( \delta_{\nodag}^{\alpha} \ve{\epsilon}_{\ve{q}\lambda}^{\beta} + \delta_{\nodag}^{\beta} \ve{\epsilon}_{\ve{q}\lambda}^{\alpha} \right) \left( 1 - e^{i \ve{q} \cdotp \ve{\delta}} \right).
\end{split}
\label{eq:dis}
\end{align}

\noindent If we define the coupling tensor

\begin{equation}
G_{\ql \ve{\delta}}^{\alpha\beta} = \frac{B_{\abs{\ve{\delta}}}^{\alpha\beta}}{2 \ve{\delta}^2} \sqrt{\frac{\hbar}{2 m \omega_{\ve{q}\lambda} N}} \left( \delta^\alpha \hat{e}^\beta_\ql + \delta^\beta \hat{e}^\alpha_\ql \right)  \left( 1 - e^{i \ve{q} \cdot \ve{\delta}} \right),
\end{equation}

\noindent the Hamiltonian may be written as

\begin{equation}
\mathcal{H}_{\rm ME} = \sum_{i,\ve{\delta}} \sum_{\alpha\beta} \sum_\ql G_{\ql \ve{\delta}}^{\alpha\beta}  S_{i}^{\alpha} S_{i+\ve{\delta}}^{\beta}  \left( c^{\dagger}_{\ql} + c^{\nodag}_{\mql} \right)  e^{i \ve{q} \cdot \ve{r}_i}.
\label{eq:HME3}
\end{equation}

We now want to expand the Hamiltonian \eqref{eq:HME3} to first order in magnon operators, as this will produce terms quadratic in the boson operators. For this expansion to be justified, we need to expand from the classical ground state of the spins. That is, we first have to express the Hamiltonian in terms of the spins in the primed coordinate system, defined in Sec. \ref{ssec:magnons}. We did a similar procedure in Sec. \ref{sssec:biaxialNiO}, where we defined $\mathbf{O}$ so that a vector in the primed coordinate system $\ve{r}'$ was related to the unprimed coordinates as $\ve{r} = \mathbf{O} \ve{r}'$. The spins $\ve{S}$ may then be written as $\ve{S} = \mathbf{O} \ve{S}'$. If we then define $\tilde{G}_{\ql \ve{\delta}}^{\alpha\beta} = (\mathbf{O}^{\rm T} \mathbf{G}_{\ql \ve{\delta}} \mathbf{O})^{\alpha\beta}$, the Hamiltonian follows as

\begin{equation}
\mathcal{H}_{\rm ME} = \sum_{i,\ve{\delta}} \sum_{\alpha\beta} \sum_\ql \tilde{G}_{\ql \ve{\delta}}^{\alpha\beta}  S_{i}'^{\alpha} S_{i+\ve{\delta}}'^{\beta}  \left( c^{\dagger}_\ql + c^{\nodag}_{\mql} \right)  e^{i \ve{q} \cdot \ve{r}_i}.
\label{eq:HME4}
\end{equation}

\noindent We are now ready to do a Holstein-Primakoff transformation of Eq. \eqref{eq:HME4}. We first split the sum over $i$ and $\ve{\delta}$ into four sums: one for each permutation of ${i, i+\ve{\delta} \in \{\mathrm{A},\mathrm{B} \}}$, where A and B are the two sublattices. We will use the following notation: ${\sum_{\ve{\delta} \in ab}}$ means that $\ve{\delta}$ is summed over vectors pointing from a site on a sublattice $a$ to all sites on a sublattice $b$. We note that $\sum_{\ve{\delta} \in \mathrm{AA}} = \sum_{\ve{\delta} \in \mathrm{BB}}$ and $\sum_{\ve{\delta} \in \mathrm{AB}} = \sum_{\ve{\delta} \in \mathrm{BA}}$ due to the equivalency of the sublattices. If we neglect terms which are of third order or higher in the boson operators, and drop the linear terms, the resulting Hamiltonian is

\begin{align}
\begin{split}
\mathcal{H}_{\rm ME} &= \sqrt{\frac{ N_{\rm A} S^3}{2}} \sum_\ql \left( c^{\dagger}_\ql + c^\nodag_\mql \right) \times \\
&\quad \Bigg\{  \left( \sum_{\ve{\delta} \in \mathrm{AA}} \tilde{G}_{\ql \ve{\delta}}^{xz} - \sum_{\ve{\delta} \in \mathrm{AB}} \tilde{G}_{\ql \ve{\delta}}^{xz} \right) \left(  a_{\ve{q}}^\nodag + a_{-\ve{q}}^\dagger  -  b_{\ve{q}}^\nodag - b_{-\ve{q}}^\dagger   \right)\\
&\: -i \left( \sum_{\ve{\delta} \in \mathrm{AA}} \tilde{G}_{\ql \ve{\delta}}^{yz} - \sum_{\ve{\delta} \in \mathrm{AB}} \tilde{G}_{\ql \ve{\delta}}^{yz} \right) \left(  a_{\ve{q}}^\nodag - a_{-\ve{q}}^\dagger  +  b_{\ve{q}}^\nodag - b_{-\ve{q}}^\dagger   \right)
 \Bigg\}.
\label{eq:me_1mag}
\end{split}
\end{align}

\noindent If we now define

\begin{equation}
\tilde{M}_{\ql} =  \sqrt{\frac{ N_{\rm A} S^3}{2}} \left[ \sum_{\ve{\delta} \in \mathrm{AA}} \left( \tilde{G}_{\ql \ve{\delta}}^{xz} - i \tilde{G}_{\ql \ve{\delta}}^{yz} \right)  - \sum_{\ve{\delta} \in \mathrm{AB}} \left( \tilde{G}_{\ql \ve{\delta}}^{xz} -
i \tilde{G}_{\ql \ve{\delta}}^{yz} \right) \right],
\label{eq:M}
\end{equation}

\noindent we may write the Hamiltonian as

\begin{equation}
\mathcal{H}_{\rm ME} = \sum_\ql \left( c^{\dagger}_\ql + c^\nodag_\mql \right) 
 \Bigg\{   \tilde{M}_{\ql}  \left(  a_{\ve{q}}^\nodag -  b_{-\ve{q}}^\dagger  \right) + \tilde{M}_{\mql}^*  \left(  a_{-\ve{q}}^\dagger - b_{\ve{q}}^\nodag   \right) \Bigg\},
\label{eq:Hmp}
\end{equation}

\noindent where we have used that $\left(\tilde{G}_{\ql \ve{\delta}}^{\alpha\beta}\right)^* = \tilde{G}_{\mql \ve{\delta}}^{\alpha\beta}$.

We have now expressed the Hamiltonian on a form where magnon and phonon operators are coupled through a single coupling coefficient $\tilde{M}_{\ql}$. All physical details of the material under consideration is contained in $\tilde{M}_{\ql}$. In its definition in Eq. \eqref{eq:M}, we summed over all neighbors on both sublattices. When performing a calculation, one naturally has to cut off this sum at some point. In NiO, the nn's of a spin site belonging to sublattice A consist of six sites belonging to sublattice A, and six sites belonging to sublattice B. All nnn's belong to sublattice B. If we generalize this to more distant neighbors, we find that for any given $n$'th layer of nearest neighbors, if $n$ is odd, then half of the neighbors belong to either sublattice. If $n$ is even, then all neighbors belong to a single sublattice. A natural choice in NiO is therefore to include nearest and next nearest neighbors in the sum over $\ve{\delta}$, as we then include one of each sort of neighbor layers. Two independent sets of magnetoelastic coefficients are therefore included, giving four coefficients in total. We define ${\tilde{B}^{\alpha\beta} = \tilde{B}^\parallel \delta^{\alpha\beta} + \tilde{B}^\perp \left(1 - \delta^{\alpha\beta} \right)}$ as the magnetoelastic coefficient in the nearest neighbor interaction, and ${B^{\alpha\beta} = B^\parallel \delta^{\alpha\beta} + B^\perp \left(1 - \delta^{\alpha\beta} \right)}$ as the magnetoelastic coefficient in the next-nearest neighbor interaction.


\section{Magnon-phonon hybridization in NiO}
\label{sec:results}
We now combine the magnon Hamiltonian \eqref{eq:Hmag}, the phonon Hamiltonian \eqref{eq:Hph} and the magnon-phonon Hamiltonian \eqref{eq:Hmp}. The full Hamiltonian can then be expressed as

\begin{equation}
\mathcal{H} = \sum_{\ve{k}} \Psi^{\rm T}_{\ve{k}} \mathbf{H}^{\nodag}_{\ve{k}} \Psi^{\nodag}_{\ve{k}},
\end{equation}

\noindent where $\Psi_{\ve{k}} = [\psi^{\nodag}_{\ve{k}},\, \psi^{\dagger}_{-\ve{k}}]^{\rm T}$ is a vector of all operators, where ${\psi_{\ve{k}} = [a^{\nodag}_{\ve{k}},\, b^{\dagger}_{-\ve{k}},\, c^{\nodag}_{1,\ve{k}},\, c^{\nodag}_{2,\ve{k}},\, c^{\nodag}_{3,\ve{k}},\, c^{\nodag}_{4,\ve{k}},\, c^{\nodag}_{5,\ve{k}},\, c^{\nodag}_{6,\ve{k}}]}$, and $\mathbf{H}_{\ve{k}}$ is a $(16 \times 16)$ non-diagonal matrix. We diagonalize $\mathbf{H}_{\ve{k}}$ with the same procedure as we did with the magnon Hamiltonian in Section \ref{ssec:magnons}. This diagonalization reveals the new hybridized eigenstates with the corresponding energy eigenvalues.

\begin{figure}[htb!]
\centering
\includegraphics[width=1.0\columnwidth]{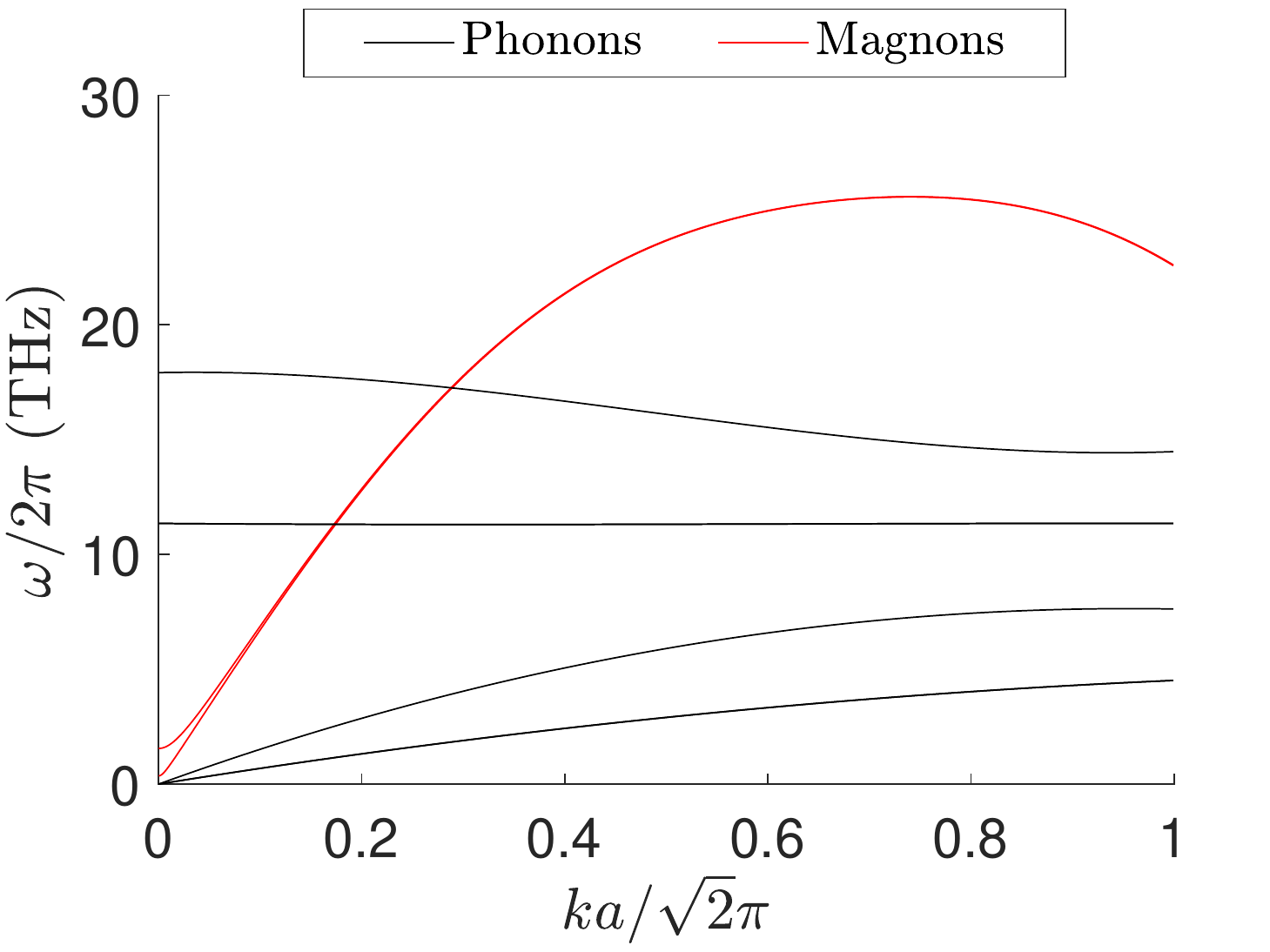}
\caption{The dispersion relation for free magnons and phonons in NiO. The black lines depict the six phonon modes, and the red lines depict the two magnon modes.}
\label{fig:disp}
\end{figure}

In order to get an initial overview of the full momentum dependence of the energy dispersions, we have plotted the energies of the free magnons and free phonons in the first phonon Brillouin zone in Fig. \ref{fig:disp}. The black lines depict the phonons, while the red lines depict the magnons. Note that the magnons are non-degenerate, which is due to the hard-axis anisotropies. As can be observed in the figure, the magnon modes cross the optical phonon modes in two distinct areas, at frequencies about $11.3$ THz and $17.3$ THz. We will refer to these areas as the first and second crossing point, respectively. These are the areas where the magnon-phonon hybridization becomes apparent, and where the modes are neither magnon- nor phonon-like. Where the hybridization is strong, the properties mix, and the modes should rather be labeled magnon-polarons. Note that the modes do not cross near the zone center, as they do in for instance YIG \cite{Kikkawa2016,Flebus2017}.

We will now include magnetoelastic coupling in the analysis. We want to display the results for realistic values of the magnetoelastic coefficients. However, to the best of our knowledge, the magnetoelastic coefficients in NiO are not precisely determined. There exist magnetostriction measurements \cite{Slack1960} which in principle may be used to determine magnetoelastic coefficients, but neither of these measurements are sufficiently detailed to determine all four coefficients we use in this analysis. We will therefore rather assess the expected order of magnitude of the coefficients. We do this by assuming that all but one coefficient are negligible, and use the magnetostriction measurements presented in Ref. \citen{Slack1960} to estimate the remaining coefficient. As a result, we find that the coefficients take values between approximately $0$ THz and $100$ THz. Given this approximative method, we do not expect the following results to be quantitatively accurate apart from the order of magnitude. However, we expect the qualitative effect of each coefficient to be accurate. Combined with the descriptive Eqs. \eqref{eq:afm_nio} and \eqref{eq:afm_nio_nontrivial}, we are able to supply a thorough analysis of the qualitative magnon-phonon hybridizations in NiO. This may in turn easily be generalized to other cubic collinear AFMs.

\begin{figure}[htb!]
\centering
\includegraphics[width=1.0\columnwidth]{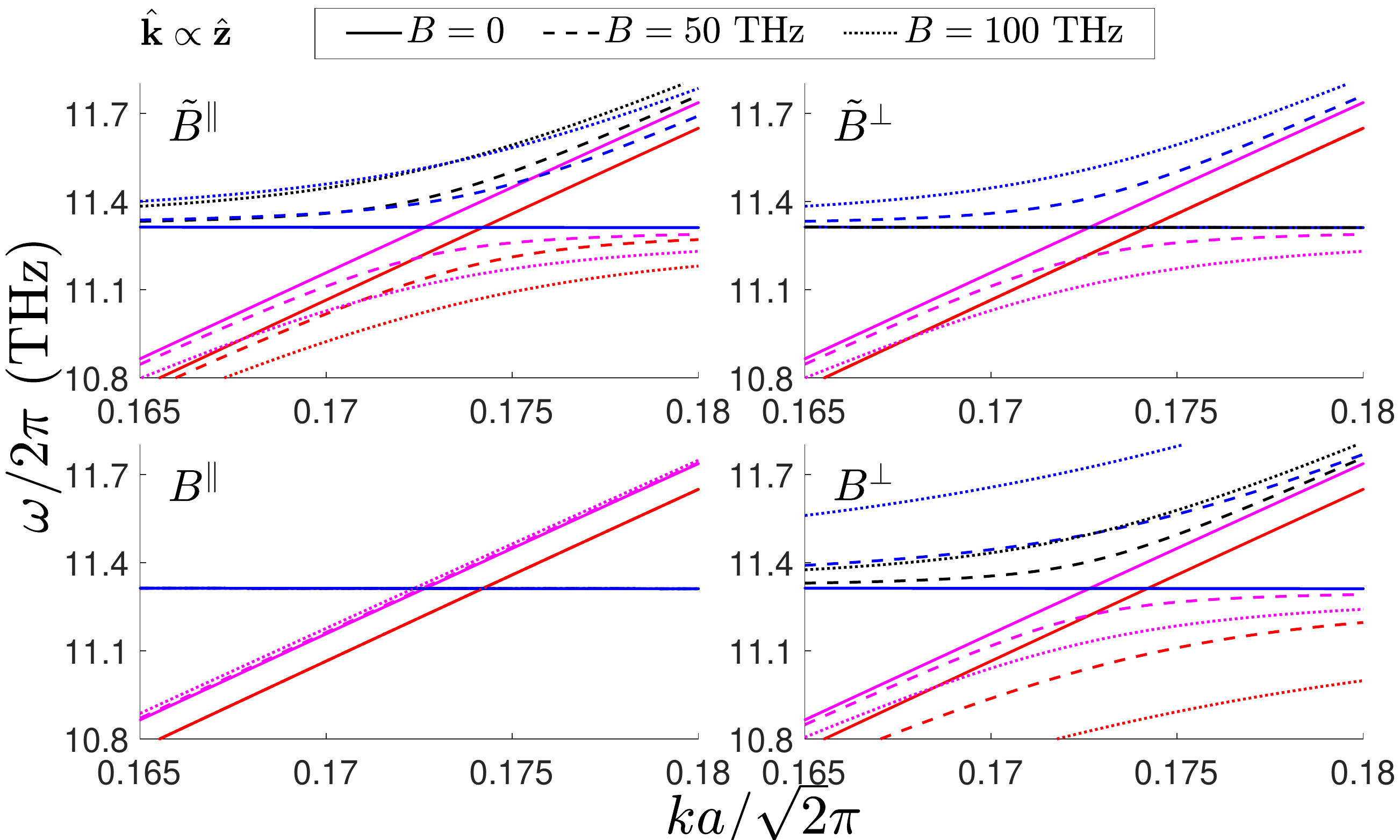}
\caption{The first crossing point, with $\ve{k} \parallel \zh$, displayed for three values of each coupling coefficient, as indicated by the legend. The non-zero coupling coefficient is given in the upper left corner of each plot, and the remaining three coefficients are set to zero.}
\label{fig:gap1z}
\end{figure}

\begin{figure}[htb!]
\centering
\includegraphics[width=1.0\columnwidth]{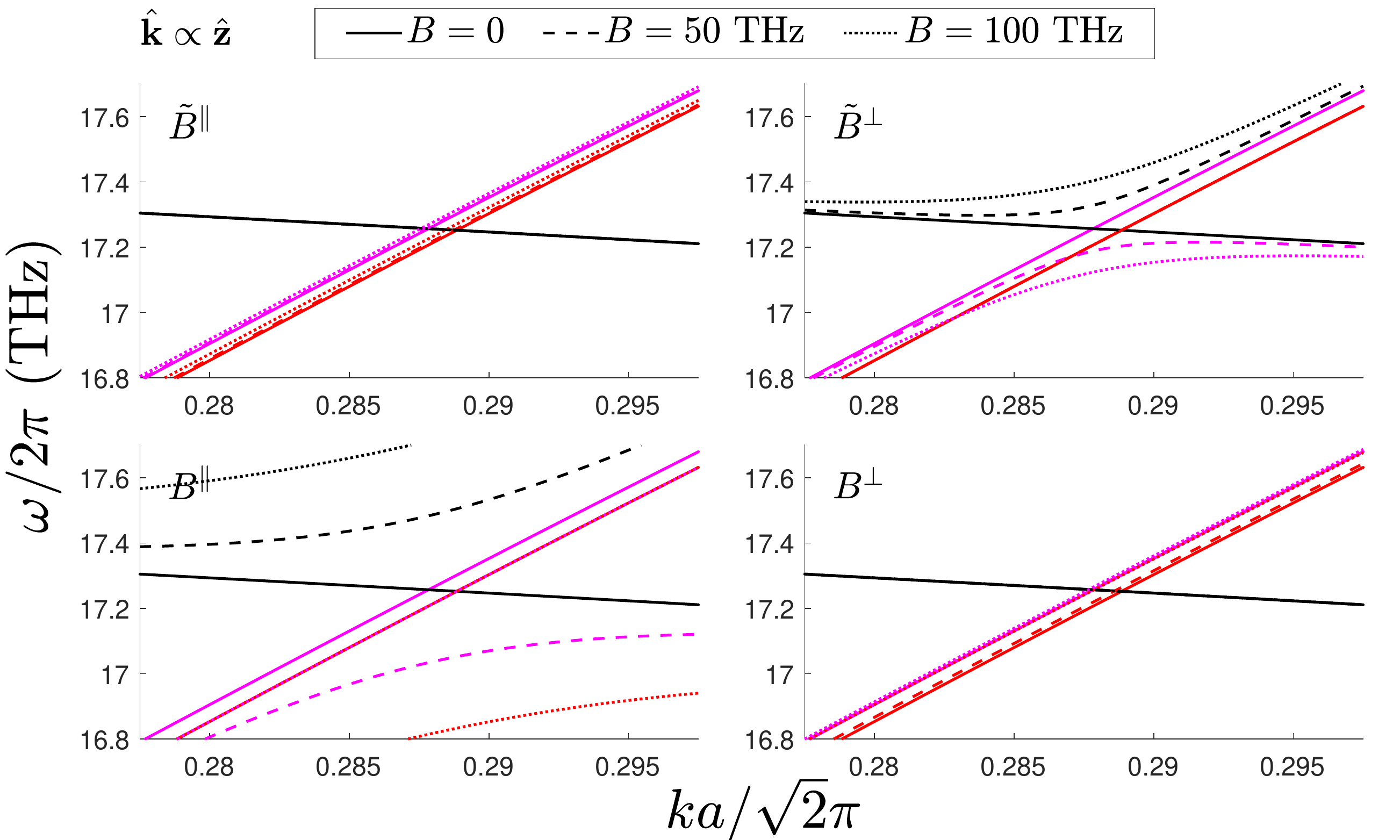}
\caption{The second crossing point, with $\ve{k} \parallel \zh$, displayed for three values of each coupling coefficient, as indicated by the legend. The non-zero coupling coefficient is given in the upper left corner of each plot, and the remaining three coefficients are set to zero.}
\label{fig:gap2z}
\end{figure}

We have plotted the magnon-phonon dispersion in the \mbox{(anti-)crossing} areas in Figs. \ref{fig:gap1z} and \ref{fig:gap2z} for different values of the magnetoelastic coefficients. As the modes now mix, the previous black/red labeling for phonons/magnons can no longer be applied. In this and all following figures, all modes are thus colored differently in order for them to be easily recognized. We have continued the assumption from the approximate assessment of the coefficients, namely that we assume that all but one coefficient are negligible, and therefore display the \mbox{(anti-)crossings} with only one non-zero coefficient at a time. We display the modes for three different non-zero values of the magnetoelastic coefficients in the range which was found to be realistic: 0, 50 and 100 THz.

Fig. \ref{fig:gap1z} displays the \mbox{(anti-)crossings} between the magnons and the TO phonon modes. There are a number of features in this plot which should be addressed. First, $B^\parallel$ causes no hybridization between the modes. This is simply due to that $B^\parallel$ only couples to longitudinal phonon modes, and thus does not affect the TO phonons. Second, $\tilde{B}^\parallel$ and $B^\perp$ apparently cause all modes to hybridize. The same conclusion can be drawn by directly reading off the hybridizations from Eqs. \eqref{eq:afm_nio} and \eqref{eq:afm_nio_nontrivial}. Third, the lower magnon mode does not couple to any TO phonons for any values of $\tilde{B}^{\perp}$. Looking at Eq. \eqref{eq:afm_nio_nontrivial}, we find that the magnon mode associated with fluctuations in $n'^y$ does not couple to the TO phonon modes if only $\tilde{B}^{\perp}$ is non-zero. As $\yh'$ is the axis with the weakest hard axis anisotropy, $n'^y$ corresponds to the lowest energy magnon, and this therefore confirms the result of Fig. \ref{fig:gap1z}.

The \mbox{(anti-)crossings} between the magnons and the LO phonon mode are displayed in Fig. \ref{fig:gap2z}. All qualitative features of this plot may be explained by analyzing Eqs. \eqref{eq:afm_nio} and \eqref{eq:afm_nio_nontrivial}. First, $\tilde{B}^\parallel$ and $B^\perp$ cause no hybridization of the modes. This follows directly from the semi-classical equations, as these coefficients do not couple to $\partial u^z/\partial r^z$. Second, $\tilde{B}^\perp$ and $B^\parallel$ makes the upper magnon mode hybridize with the LO phonon, also in line with the predictions of Eqs. \eqref{eq:afm_nio} and \eqref{eq:afm_nio_nontrivial}. Third, the lower magnon mode do not couple to the longitudinal phonon mode at all. Having recognized the lower magnon mode as oscillations in $n'^y$, this result was also implied by the semi-classical analysis.

\begin{figure}[htb!]
\centering
\includegraphics[width=1.0\columnwidth]{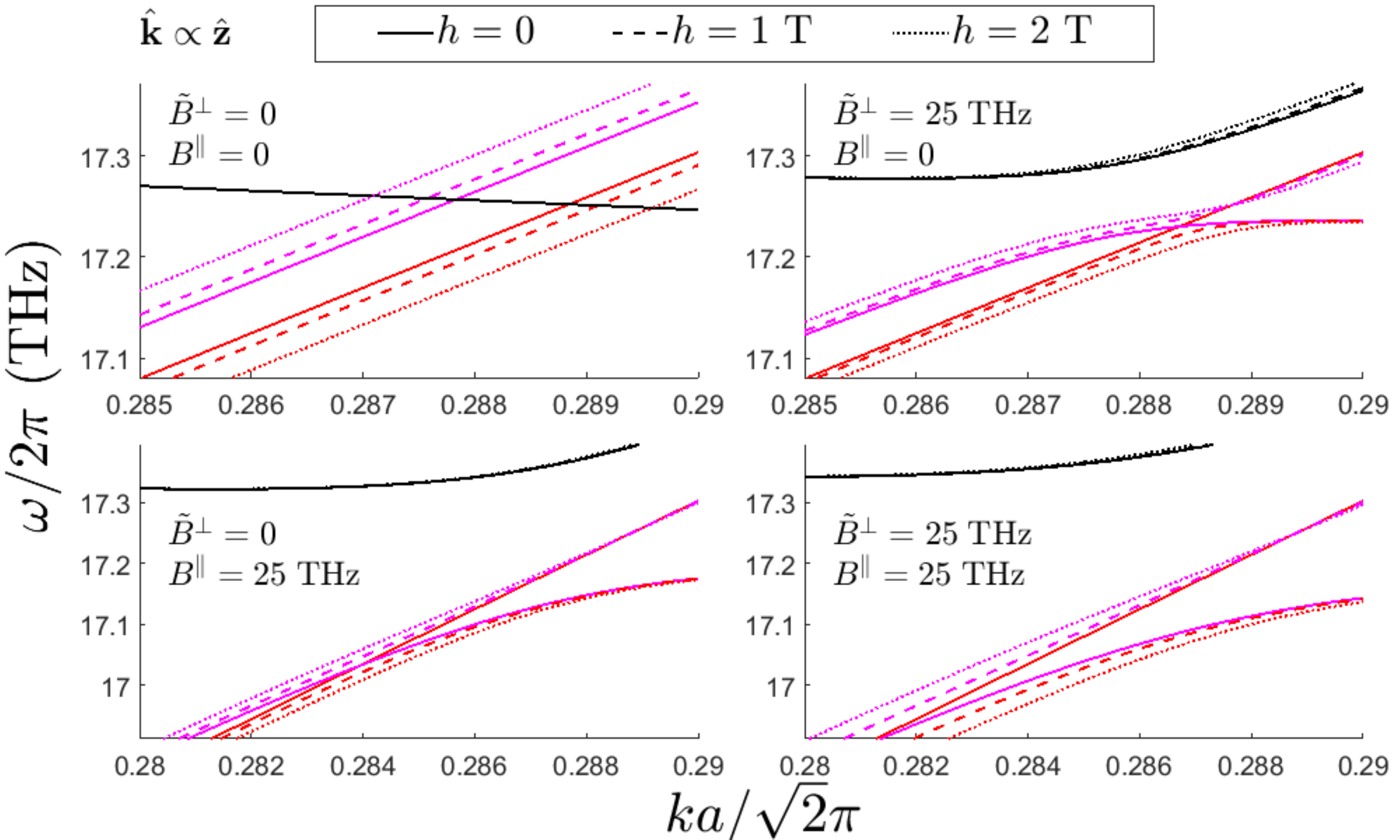}
\caption{A zoomed-in plot of the second crossing point for three different magnetic field strengths, and with different values of $\tilde{B}^\perp$ and $B^\parallel$. In all instances,  $\ve{k} \propto \zh$, $\ve{h} = h \zh'$ and $\tilde{B}^\parallel = B^\perp = 0$. In the absence of an external magnetic field, the lower magnon mode is decoupled from the longitudinal optical phonon. When an external magnetic field is applied, this magnon mode hybridizes with the magnon-polaron mode if $\tilde{B}^\perp$ and/or $B^\parallel$ are finite.}
\label{fig:hgap2z}
\end{figure}

Another prediction of the semi-classical analysis in Secs. \ref{sssec:biaxialNiO} and \ref{sssec:biaxialNiO_nontrivial} was that turning on a magnetic field would lift the decoupling of the LO phonon mode and the lower magnon mode. We therefore supply additional plots of the \mbox{(anti-)crossings} between the magnon modes and the LO phonon mode, this time with an external magnetic field present along the $\zh'$ axis, in Fig. \ref{fig:hgap2z}. We do this for three different magnetic field strengths, 0, 1 and 2 T, well below the spin-flop field where the quantum theory is expected to be imprecise to this order in the magnon operators \cite{Machado2017}. The four plots display the results for the four permutations of $\tilde{B}^\parallel, B^\perp \in \{ 0,\, 25 \}$ THz. The plots confirm that applying a magnetic field couples all modes. Increasing the magnetic field strength shows hybridization between the previosuly uncoupled modes, which confirms that the hybridization is tunable. We have limited the magnetoelastic coefficients to 25 THz simply because this gives more readable plots. Larger $B^\parallel$ causes the hybridization to be very strong, as displayed in the lower left plot of Fig. \ref{fig:gap2z}. This causes the upper magnon mode to cross the lower magnon mode at lower $k$, and the effect of applying a magnetic field would hence be most evident at lower $k$, making the plots somewhat less coherent. This effect can be seen in the two lower plots, where $B^\parallel$ is non-zero. The qualitative effect of applying a magnetic field is nonetheless also present for larger values of the magnetoelastic coefficients.

\section{Discussion}
\label{sec:discussion}
The quantitative analysis in the previous section has provided results for the magnon-phonon hybridizations in NiO given a selection of values for the magnetoelastic coefficients. We have furthermore found that the semi-classical analysis given in Secs. \ref{sssec:biaxialNiO} and \ref{sssec:biaxialNiO_nontrivial} is able to describe all of these hybridizations qualitatively. Both approaches are fairly general, and the methods can straightforwardly be applied to other cubic collinear AFMs. The specific results for NiO can also to some extent be generalized to other cubic collinear antiferromagnetic systems, at least qualitatively.

Antiferromagnetic cubic materials introduce at least two new features to magnon-polarons as compared to their ferromagnetic counterparts. The first feature is that longitudinal phonon modes hybridize with magnons in many antiferromagnetic structures. There are two origins of this hybridization: first, the longitudinal modes hybridize if the anisotropies cause the spins to condense non-parallel to any of the crystal axes; second, this occurs due to the second term in the magnetoelastic Hamiltonian \eqref{eq:Hafm_nontrivial}, which is there due to non-trivial spin ordering in the ground state. The first origin is not unique to AFMs. Indeed, the spins condense in the [111] directions in ferromagnets such as pure nickel and magnetite \cite{Bozorth1950,Medrano1999}. One should expect the longitudinal phonons propagating along the crystal axes to hybridize with magnons in these ferromagnetic materials as well. This follows directly from a ferromagnetic analogy of Eq. \eqref{eq:afm_ellipse}, where the Néel field is substituted with magnetization and where one of the magnon polarizations is discarded (the evanescent mode). The second origin however, is only attainable if there are at least two spin sublattices involved, as this is the least requirement for creating a non-trivial spin-structure. This latter effect introduce hybridization between longitudinal phonons and magnons even if the spins are aligned along one of the crystal axes in the ground state.

The second feature we have observed is that the hybridization between antiferromagnetic magnons and the phonons is tunable by an external magnetic field. Moreover, the hybridization may be switched off entirely in certain structures, and thus shows potential for binary control. This is a unique feature of bipartite magnetic structures. Although magnon-phonon hybridizations in ferromagnets can be tuned in the sense that the magnon energy can be increased relative to the phonon modes, the interaction in itself cannot be completely turned on and off. This may however be done in certain AFMs, where NiO is a prominent example of a material in which this phenomenon should occur. This is due to there being two magnon modes in AFMs, which enables "moving" the magnetoelastic interaction between the magnon modes by changing the properties of the eigenstates. More precisely, since applying a magnetic field changes the spin oscillations of the magnon eigenmodes qualitatively, and because the hybridizations is sensitive to this oscillation pattern, we may use the magnetic field to tune the hybridization in AFMs.

We expect magnon-polarons in AFMs to be found at optical phonon energies, which are usually well up in the THz range. In contrast, magnon-polarons in ferromagnets are typically found at the crossings between magnons and acoustic phonons. This is due to the linear dispersion of antiferromagnetic magnons at low $k$. The effect of this is that the magnon dispersions never cross the acoustic phonon dispersions if their velocity is greater than that of the phonons. Followingly, there is no strong hybridization between the magnons and acoustic phonons. The high frequencies at which magnons-polarons are found in AFMs make the accessibility of antiferromagnetic magnon-polarons lower than the ferromagnetic ones. This in turn might reduce their applicability. For instance, magnon-polarons have shown to play an important role in the SSE effect in YIG \cite{Kikkawa2016,Flebus2017}, as the properties of phonons affect the induced spin current. This would not occur in most AFMs, as the magnon-polarons are not necessarily thermally accessible below the Néel temperature of the material.

An important point to address is, precisely, how to access antiferromagnetic magnon-polarons. Both crossings between magnon-like and phonon-like modes in NiO occur far away from the zone center. This stands in contrast to ferromagnetic magnon-polarons, which are typically found at low $k$. For that reason, magnon-polarons are not accessible by for instance conventional first order Raman scattering, which aims to excite modes at very long wavelengths due to the negligible momentum of the photon. We have already discarded thermal excitation as an alternative, due to the high energies. The magnon-polarons may be accessed with neutron scattering, which previously has been used to map the dispersion relations of both magnons and phonons separately in NiO. The most promising way of accessing the magnon-polarons in AFMs might however be with femtosecond optics \cite{Kampfrath2010,JuPRL2004,BattiatoPRL2010,BerrittaPRL2016}.

Injection of coherent phonons at high frequencies in the THz range has been achieved using ultrafast lasers~\cite{Ruello2015}. In a direct analogy with spin pumping driven via coherent elastic waves~\cite{Weiler2012}, and subsequently formed magnon-polarons~\cite{Kamra2015,Hayashi2018}, in ferromagnets, these coherent THz phonons may directly excite the corresponding magnon-polarons, and result in a spin pumping current~\cite{Kamra2015,Hayashi2018}. The latter may be detected electrically via inverse spin Hall effect (ISHE). Since a magnetic field may be used to tune the hybridizations, one might even be able to tune the measured ISHE voltage with the magnetic field.


\section{Conclusion}
\label{sec:conclusion}
We explore magnon-polarons in cubic collinear antiferromagnets, and focus on their qualitative difference to their ferromagnetic counterparts. We find that there are two distinct features of antiferromagnetic magnon-polarons. First, antiferromagnetic materials with either a complex spin structure or spin alignment axis non-parallel to any of the cubic axes generally result in hybridization between magnons and longitudinal phonons. Second, the hybridizations may be tuned by an external magnetic field by changing the qualitative properties of the magnons eigenmodes. NiO is an example of such a material, where a hybridization may even be turned on and off.

Magnons-polarons in antiferromagnets seem to have a reduced applicability compared to ferromagnetic ones due to their high energies and locations at intermediate momenta. However, the rapid evolving fields of ultrafast dynamics and femtosecond optics provide tools for an easier access to the antiferromagnetic magnon-polarons. We suggest spin pumping driven via coherently injected phonons as a promising approach for investigating these magnon-polarons.

\section*{Acknowledgements}
This work was supported by the Research Council of
Norway through its Centres of Excellence funding scheme,
Project No. 262633 "QuSpin", as well as by the European Research
Council via Advanced Grant No. 669442 "Insulatronics".

\appendix

\section{Derivation of the magnetoelastic Hamiltonian}
\label{sec:app1}

Magnetoelastic coupling relates magnetic and elastic degrees of freedom. In a discrete lattice, at a site $i$, the magnetic moment is proportional to the spin $\ve{S}_i$.  A measure of the elastic degree of freedom is the displacement of atom $i$, $\ve{u}_i$. The magnetoelastic coupling depends on the spin $\ve{S}_i$ and displacement $\ve{u}_j$ at all sites $i$ and $j$. We will now discuss phenomenological models of the magnetoelastic coupling starting with the simplest possible forms. We consider materials in which the magnetic atoms form a Bravais lattice, all of which are invariant under inversion symmetry. 

Let us begin by discussing a simple ansatz, that the discrete magnetoelastic model is bilinear in spin and displacement,

\begin{equation}
\mathcal{H}_{\rm ME}^{\rm I} = \sum_{ij} \sum_{\alpha\beta} B_{ij}^{\alpha\beta} S_i^{\alpha} u_j^{\beta} \,.
\label{eq:Hsimple1}
\end{equation}

\noindent Here, $B_{ij}^{\alpha\beta}$ is a phenomenological coupling tensor relating the spin at site $i$ with the atom displacement at site $j$. Neumann's principle states that the physical properties of a crystal must share the symmetries of the crystal \cite{Birss1964,KATZIR2004}. Under a transformation $\mathcal{R}$, the spin transforms as ${\ve{S} \rightarrow \abs{\mathcal{R}} \mathcal{R} \ve{S}}$ and the displacement transforms as ${\ve{u} \rightarrow \mathcal{R} \ve{u}}$. Since we have assumed that the magnetic atoms form a Bravais lattice, the Hamiltonian is invariant under the inversion operation $\mathcal{R} = -1$ and $\abs{\mathcal{R}} = -1$. Consequently, $B_{ij}^{\alpha\beta} = -B_{ij}^{\alpha\beta}$. The only solution is $B_{ij}^{\alpha\beta} = 0$. Hence, the Hamiltonian \eqref{eq:Hsimple1} does not contribute.

Let us proceed by first restoring invariance under the inversion operation without expanding to higher orders in the magnetic or elastic degrees of freedom. One may observe that introducing another quantity that transforms as a vector fulfills the requirement of invariance. In our system, the only natural vector we have left is the position vector $\ve{r}_i$. Our next attempt is therefore

\begin{equation}
\mathcal{H}_{\rm ME}^{\rm II} = \sum_{ijk} \sum_{\alpha\beta\gamma} B_{ijk}^{\alpha\beta\gamma} S_i^{\alpha} u_j^{\beta} r_k^\gamma \, ,
\label{eq:Hsimple2}
\end{equation}

\noindent which will have finite elements $B_{ijk}^{\alpha\beta\gamma}$ even when requiring inversion invariance.

There are also other constraints. The Hamiltonian (\ref{eq:Hsimple2}) must be invariant under uniform translations of the lattice. Mathematically, this can be expressed by a uniform displacement $u_j^\beta \rightarrow u_j^\beta + \delta u^\beta$, or by a uniform shift of the position vectors $r_k^\gamma \rightarrow r_k^\gamma + \delta r^\gamma$. The Hamiltonian (\ref{eq:Hsimple2}) must be invariant under both of these transformations separately. 

We begin by considering a uniform displacement, where the invariance requirement reads

\begin{equation}
\sum_{ijk} \sum_{\alpha\beta\gamma} B_{ijk}^{\alpha\beta\gamma} S_i^{\alpha} \delta u^{\beta} r_k^\gamma = 0.
\end{equation}

\noindent This relation must hold for all different spin configurations $\{S_i^\alpha\}$ and atom configurations $\{ r_k^\gamma \}$. Furthermore, as $\delta u^\beta$ is arbitrary, the relation must hold for every component $\beta$. The resulting constraint for the magnetoelastic coefficients is therefore

\begin{equation}
\sum_{j} B_{ijk}^{\alpha\beta\gamma} = 0.
\label{eq:constraint1}
\end{equation}

We now make use of the second invariance requirement. A uniform shift of the position vectors $r_k^\gamma \rightarrow r_k^\gamma + \delta r^\gamma$ should leave the Hamiltonian unchanged:

\begin{equation}
\sum_{ijk} \sum_{\alpha\beta\gamma} B_{ijk}^{\alpha\beta\gamma} S_i^\alpha u_j^\beta \delta r^\gamma = 0.
\end{equation}

\noindent This relation must hold for any spin configuration $\{ S_i^\alpha \}$ and atom displacements $\{ u_j^\beta \}$. As the shift $\delta r^\gamma$ is arbitrary, the relation must hold for every component $\gamma$. Thus, we are left with the constraint

\begin{equation}
\sum_{k} B_{ijk}^{\alpha\beta\gamma} = 0.
\label{eq:constraint2}
\end{equation}

\noindent Requiring the Hamiltonian to be translationally invariant induces two constraints on the tensor $B_{ijk}^{\alpha\beta\gamma}$, Eqs. \eqref{eq:constraint1} and \eqref{eq:constraint2}.

Let us now inspect the Hamiltonian \eqref{eq:Hsimple2} more closely. $S_i^\alpha$ and $u_j^\beta$ are dynamical variables. In contrast, $r_k^\gamma$ is fixed when the lattice properties are defined. In other words, $\{ r_k^\gamma \}$ defines the equilibrium lattice, and the dynamics related to displacements from equilibrium is contained in $\{ u_j^\beta \}$. Keeping this in mind, we could define an effective coupling tensor as ${\tilde{B}_{ij}^{\alpha\beta} = \sum_k \sum_\gamma B_{ijk}^{\alpha\beta\gamma} r_k^\gamma}$, so that the Hamiltonian \eqref{eq:Hsimple2} reads

\begin{equation}
\mathcal{H}^{\rm II}_{\rm ME} = \sum_{ij} \sum_{\alpha\beta} \tilde{B}_{ij}^{\alpha\beta} S_i^\alpha u_j^\beta.
\label{eq:Hsimple2_1}
\end{equation}

\noindent At first sight Eq.\ (\ref{eq:Hsimple2_1}) might seem to have become analogous to the starting ansatz \eqref{eq:Hsimple1}, the latter of which does not contribute. However, there is an important distinction since $\tilde{B}_{ij}^{\alpha\beta}$ is not a tensor of constant coefficients. Instead, $\tilde{B}_{ij}^{\alpha\beta}$ is a sum of products between a tensor of constant coefficients and position vector components. Therefore, the transformation properties of $\tilde{B}_{ij}^{\alpha\beta}$ differ from the ones of $B_{ij}^{\alpha\beta}$. As a result, the Hamiltonian \eqref{eq:Hsimple2_1} essentially differs from the starting Hamiltonian \eqref{eq:Hsimple1}. 

By expressing the Hamiltonian \eqref{eq:Hsimple2} as in \eqref{eq:Hsimple2_1}, we realize that several $k$ and $\gamma$ components of $B_{ijk}^{\alpha\beta\gamma}$ contribute to the effective tensor $\tilde{B}_{ij}^{\alpha\beta}$, but the relative contribution to the sum is not important. The only physical significance of the introduction of the position $r_k^\gamma$ is its transformation properties. This implies that, without a loss of generality, we can choose a selection of the tensor elements $B_{ijk}^{\alpha\beta\gamma}$ to be equal to zero as long as we do not break any symmetries of the lattice under consideration. There is an infinite number of such choices in an infinite lattice.  We will follow a path that is physically transparent because, in the continum limit, it couples the spins to strain tensor components. As will be evident, we obtain this by leaving $B_{ijk}^{\alpha\beta\gamma}$ finite for the following indices $i$, $j$, and $k$: if $i \neq j$ then $k \in \{ i,j \}$, and if $i=j$ then $k$ can point to any lattice site. $B_{ijk}^{\alpha\beta\gamma}$ is set to zero for all other $k$'s. 

We may now insert the definition of the non-zero tensor elements $B_{ijk}^{\alpha\beta\gamma}$ into the constraints \eqref{eq:constraint1} and \eqref{eq:constraint2}. The constraint \eqref{eq:constraint1} then reads

\begin{align}
B_{iij}^{\alpha\beta\gamma} + B_{ijj}^{\alpha\beta\gamma} &= 0 
\label{eq:cstr1}, \\
B_{iii}^{\alpha\beta\gamma} &= - \sum_{k \neq i} B_{iki}^{\alpha\beta\gamma} 
\label{eq:cstr4}.
\end{align}

\noindent The second constraint \eqref{eq:constraint2} reads

\begin{align}
B_{iji}^{\alpha\beta\gamma} + B_{ijj}^{\alpha\beta\gamma} &= 0 
\label{eq:cstr2}, \\
B_{iii}^{\alpha\beta\gamma} &= - \sum_{k \neq i} B_{iik}^{\alpha\beta\gamma} 
\label{eq:cstr3}.
\end{align}

\noindent In all of these equations, $i \neq j$. Eqs. \eqref{eq:cstr1} and \eqref{eq:cstr2} imply ${B_{iij}^{\alpha\beta\gamma} = B_{iji}^{\alpha\beta\gamma} = -B_{ijj}^{\alpha\beta\gamma}}$. We therefore define a new tensor ${B_{ij}^{\alpha\beta\gamma} \equiv B_{ijj}^{\alpha\beta\gamma}}$, and insert all the constraints above into the Hamiltonian \eqref{eq:Hsimple2}. The resulting Hamiltonian is

\begin{equation}
\mathcal{H}^{\rm II}_{\rm ME} = \sum_{ij} \sum_{\alpha\beta\gamma} B_{ij}^{\alpha\beta\gamma} S_i^{\alpha} \left( u_i^{\beta} - u_{j}^{\beta} \right) \left( r_i^\gamma - r_j^\gamma \right),
\label{eq:Hsimple2_2}
\end{equation}

\noindent where both $i$ and $j$ run over all lattice sites. This Hamiltonian \eqref{eq:Hsimple2_2} is thus the lowest order non-zero magnetoelastic Hamiltonian. 

The Hamiltonian \eqref{eq:Hsimple2_2} describes both rotations, which are antisymmetric in $\beta\gamma$, and strains, which are symmetric in $\beta\gamma$. We can separate these two effects by defining the matrices

\begin{align}
R^{\beta\gamma}_{\mathrm{S}, ij} &= \frac{1}{2} \frac{1}{| \ve{r}_i - \ve{r}_j |^2} \Bigg[ \left( r_i^{\gamma} - r_j^{\gamma} \right)\left( u_i^{\delta} - u_j^{\delta} \right) + \left( r_i^{\delta} - r_j^{\delta} \right)\left( u_i^{\gamma} - u_j^{\gamma} \right) \Bigg],
\label{eq:disc_strain1} \\
R^{\beta\gamma}_{\mathrm{R}, ij} &= \frac{1}{2} \frac{1}{| \ve{r}_i - \ve{r}_j |^2} \Bigg[ \left( r_i^{\gamma} - r_j^{\gamma} \right)\left( u_i^{\delta} - u_j^{\delta} \right) - \left( r_i^{\delta} - r_j^{\delta} \right)\left( u_i^{\gamma} - u_j^{\gamma} \right) \Bigg],
\label{eq:disc_rot1}
\end{align}

\noindent where $R^{\beta\gamma}_{\mathrm{S}, ij}$ captures strains and $R^{\beta\gamma}_{\mathrm{R}, ij}$ captures rotations. In the following, we will restrict the analysis to strains only, and we therefore drop the rotation term. We name $R^{\beta\gamma}_{\mathrm{S}, ij}$ the discrete strain tensor, and denote it simply as $R^{\beta\gamma}_{ij}$ from now on. By summing over the vector $\ve{\delta} \equiv \ve{r}_j - \ve{r}_i$ instead of $j$, we may rewrite the Hamiltonian as

\begin{equation}
\mathcal{H}^{\rm II}_{\rm ME} = \sum_{i, \ve{\delta}} \sum_{\alpha\beta\gamma} B_{|\ve{\delta}|}^{\alpha\beta\gamma} S_i^{\alpha} R_{i,i+\ve{\delta}}^{\beta\gamma},
\label{eq:Hsimple2_3}
\end{equation}

\noindent The $ij$-index of $B_{ij}^{\alpha\beta\gamma}$ was changed to $|\ve{\delta}|$. This is possible due to translational and rotational invariance of the Bravais lattice. Due to the normalization factor $| \ve{r}_i - \ve{r}_j |^{-2}$ in the definition of the strain tensor \eqref{eq:disc_strain1}, the coupling tensor elements appearing in \eqref{eq:Hsimple2_3} are related to the tensor elements in \eqref{eq:Hsimple2_2} by the inverse of this normalization factor. As a result, $B_0^{\alpha\beta\gamma} = 0$, and the first non-zero coefficients appear in the nearest neighbor interaction.

We may now use Neumann's principle to derive the selection rules of $B_{|\ve{\delta}|}^{\alpha\beta\gamma}$. We find that there are 18 independent coefficients in triclinic crystals, 8 in monoclinic crystals, and fewer as we increase the symmetry. In cubic crystals, we find that $B_{|\ve{\delta}|}^{\alpha\beta\gamma} = 0$ is the only solution. We conclude that the Eq.\ (\ref{eq:Hsimple2_3}) cannot describe any magnetoelastic coupling in cubic crystals.

As we want the magnetoelastic Hamiltonian to be able to describe cubic antiferromagnets, we must include an additional term. Our next attempt is quadratic in the spin degrees of freedom,

\begin{equation}
\mathcal{H}_{\rm ME}^{\rm III} = \sum_{ijkl} \sum_{\alpha\beta\gamma\lambda} B_{ijkl}^{\alpha\beta\gamma\lambda} S_{i}^{\alpha} S_{j}^{\beta} u_k^\gamma r_l^\lambda.
\label{eq:H5}
\end{equation}

\noindent Just as we did for $\mathcal{H}^{\rm II}_{\rm ME}$, we require the Hamiltonian to be invariant under uniform translations of the lattice, that is ${u_k^\gamma \rightarrow u_k^\gamma + \delta u^\gamma}$ and ${r_l^\lambda \rightarrow r_l^\lambda + \delta r^\lambda}$, and obtain the constraints

\begin{align}
\sum_k B_{ijkl}^{\alpha\beta\gamma\lambda} &= 0,
\label{eq:cstr3_1} \\
\sum_l B_{ijkl}^{\alpha\beta\gamma\lambda} &= 0.
\label{eq:cstr3_2}
\end{align}
In the following, we will use a local approximation. We assume that the interaction between $\ve{S}_i$, $\ve{S}_j$ and $\ve{u}_k$ is dominated by the terms where $k \in \{ i,j \}$. Intuitively, this follows if we view the Hamiltonian (\ref{eq:H5}) as a distance dependent exchange interaction. Hence, the relevant displacements are the displacements of the involved spins. In other words, we can view it as a local expansion in the lattice distortions around the spins.

As above, where we discovered an arbitrariness in the indices of the coupling tensor related to the position vector $r_l^\lambda$, we may choose to set  $B_{ijkl}^{\alpha\beta\gamma\lambda} = 0$ for a selection of $l$'s without a loss of generality. We would like the Hamiltonian (\ref{eq:H5}) to be consistent with the continuum limit result of spins coupling to the strain tensor. We therefore choose $B_{ijkl}^{\alpha\beta\gamma\lambda}$ to be non-zero only if $l \in \{ i,j \}$, in addition to the already mentioned $k \in \{ i,j \}$. We now insert the choice of non-zero tensor elements into the constraints of Eqs.\ (\ref{eq:cstr3_1}) and (\ref{eq:cstr3_2}) derived above. The first constraint \eqref{eq:cstr3_1} then reads

\begin{align}
B_{ijii}^{\alpha\beta\gamma\lambda} + B_{ijji}^{\alpha\beta\gamma\lambda} &= 0,
\label{eq:cstr3_3} \\
B_{ijij}^{\alpha\beta\gamma\lambda} + B_{ijjj}^{\alpha\beta\gamma\lambda} &= 0,
\label{eq:cstr3_4} \\
B_{iiii}^{\alpha\beta\gamma\lambda} &= 0.
\label{eq:cstr3_5}
\end{align}

\noindent Additionally, the second constraint \eqref{eq:cstr3_2} reads
 
 \begin{align}
B_{ijii}^{\alpha\beta\gamma\lambda} + B_{ijij}^{\alpha\beta\gamma\lambda} &= 0,
\label{eq:cstr3_6} \\
B_{ijji}^{\alpha\beta\gamma\lambda} + B_{ijjj}^{\alpha\beta\gamma\lambda} &= 0.
\label{eq:cstr3_7}
\end{align}

\noindent In all constraints, $i \neq j$. The constraints \eqref{eq:cstr3_3} and \eqref{eq:cstr3_6} together imply $B_{ijii}^{\alpha\beta\gamma\lambda} = -B_{ijij}^{\alpha\beta\gamma\lambda} = -B_{ijji}^{\alpha\beta\gamma\lambda}$, and the constraints \eqref{eq:cstr3_4} and \eqref{eq:cstr3_7} additionally imply $B_{ijii}^{\alpha\beta\gamma\lambda} = B_{ijjj}^{\alpha\beta\gamma\lambda}$. Let us therefore define $B_{ij}^{\alpha\beta\gamma\lambda} \equiv B_{ijii}^{\alpha\beta\gamma\lambda}$. The resulting Hamiltonian follows as

\begin{equation}
\mathcal{H}_{\rm ME}^{\rm III} = \sum_{i \neq j} \sum_{\alpha\beta\gamma\lambda} B_{ij}^{\alpha\beta\gamma\lambda} S_{i}^{\alpha} S_{j}^{\beta} \left( u_i^{\gamma} - u_{j}^{\gamma} \right) \left( r_i^\lambda - r_j^\lambda \right),
\label{eq:Hsimple3_1}
\end{equation}

\noindent If we consider strains only, and disregard rotations, the Hamiltonian reads

\begin{equation}
\mathcal{H}_{\rm ME}^{\rm III} = \sum_{i,\ve{\delta}} \sum_{\alpha\beta\gamma\lambda} B_{|\ve{\delta}|}^{\alpha\beta\gamma\lambda} S_{i}^{\alpha} S_{i+\ve{\delta}}^{\beta} R_{i,i+\ve{\delta}}^{\gamma\lambda},
\label{eq:Hsimple3_2}
\end{equation}

\noindent where we changed the summation variable $j$ to $\ve{\delta} = \ve{r}_j - \ve{r}_i$ just as we did in the derivation of \eqref{eq:Hsimple2_3}. $\ve{\delta}$ runs over all lattice sites except $\ve{\delta} = 0$, fulfilling constraint \eqref{eq:cstr3_5}. 

One may verify that the Hamiltonian \eqref{eq:Hsimple3_2} has non-zero contributions both for cubic crystals and for crystals subject to a uniform strain. We conclude that the Hamiltonians in Eqs. \eqref{eq:Hsimple2_3} and \eqref{eq:Hsimple3_2} combined give the lowest order phenomenological picture of magnetoelastic coupling in magnetic crystals forming a Bravais lattice.

\section{Physical origin of the magnetoelastic Hamiltonians}
\label{sec:app2}

The purpose of this section is to discuss the physical origins of the magnetoelastic Hamiltonians (\ref{eq:Hsimple2_3}) and (\ref{eq:Hsimple3_2}) that we derived in Appendix\ \ref{sec:app1}. To elucidate the properties, we use as a starting point well-known interactions, and show how these generate Eqs.\ (\ref{eq:Hsimple2_3}) and (\ref{eq:Hsimple3_2}). First, we consider spin-orbit interaction between a spin and the orbit of its neighboring ions and demonstrate that this leads to a Hamiltonian equivalent to \eqref{eq:Hsimple2_3}. Second, we consider a distance dependent exchange interaction and observe that this reproduces the Hamiltonian \eqref{eq:Hsimple3_2}.

Consider first the spin-orbit coupling between a spin at site $i$ and the orbital magnetic momentum of an ion at site $j$. The magnetic moment of the spin is $\ve{\mu}^{\rm spin}_i = \gamma_{\rm s} \ve{S}_i$, where $\gamma_{\rm s}$ is the gyromagnetic ratio of the spin. The orbital magnetic moment of ion $j$ in the rest frame of spin $\ve{S}_i$ is $\ve{\mu}^{\rm ion}_j = \gamma_{\rm ion} \ve{L}_j$, where $\gamma_{\rm ion}$ is the gyromagnetic ratio of the ion, and $\ve{L}_j$ is its orbital angular momentum. The orbital angular momentum is ${\ve{L}_j = m (\ve{r}_i - \ve{r}_j) \cross \partial_t (\ve{u}_i - \ve{u}_j) }$, where $m$ is the mass of the ion and $\partial_t = \partial/ (\partial t)$ is the time differential operator. A general form of the corresponding spin-orbit Hamiltonian follows as

\begin{equation}
\mathcal{H}_{\rm SOC} = \sum_{ij} C_{ij}^{\alpha\beta\gamma} S_i^\alpha (r_i^\beta - r_j^\beta) \partial_t (u_i^\gamma - u_j^\gamma),
\end{equation}

\noindent where all constants are contained in the coupling tensor $C_{ij}^{\alpha\beta\gamma}$. Assuming plane wave solutions of the displacements, ${u_i^\gamma = U_{\ve{k}}^\gamma \exp{i \ve{k} \cdot \ve{r}_i - i \omega t}}$, where $\ve{k}$ is the wave vector of the plane wave, gives $\partial_t (u_i^\gamma - u_j^\gamma) = -i \omega (u_i^\gamma - u_j^\gamma) $. The resulting spin-orbit Hamiltonian becomes

\begin{equation}
\mathcal{H}_{\rm SOC} = \sum_{ij} \tilde{C}_{ij}^{\alpha\beta\gamma}(\omega) S_i^\alpha (r_i^\beta - r_j^\beta) (u_i^\gamma - u_j^\gamma),
\label{eq:Hsoc}
\end{equation}

\noindent where $\tilde{C}_{ij}^{\alpha\beta\gamma}(\omega) =  -i \omega C_{ij}^{\alpha\beta\gamma}$. We recognize that the expression (\ref{eq:Hsoc}) is equivalent to the Hamiltonian \eqref{eq:Hsimple2_3} after symmetrization in $\beta\gamma$.

We next consider a general exchange interaction (not excluding Dzyaloshinskii-Moriya interaction),

\begin{equation}
\mathcal{H}_{s} = \sum_{ij} \sum_{\alpha\beta} A_{ij}^{\alpha\beta} S_i^\alpha S_j^\beta,
\label{eq:H3origin}
\end{equation}

\noindent where $A_{ij}^{\alpha\beta}$ is a tensor of coefficients coupling spin components $\alpha$ and $\beta$ at lattice sites $i$ and $j$. $A_{ij}^{\alpha\beta}$ depends on the distance separating the spins $\ve{S}_i$ and $\ve{S}_j$. Let $\ve{r}_i - \ve{r}_j$ be the equilibrium position vector separating the spins, and let $\ve{u}_i - \ve{u}_j$ be the displacement relative to equilibrium. We may then expand $A_{ij}^{\alpha\beta}$ to first order in the relative displacement as

\begin{equation}
A_{ij}^{\alpha\beta}(\ve{u}_i - \ve{u}_j) \approx A_{ij}^{\alpha\beta}(0) + \sum_{\gamma} \left. \frac{\partial A_{ij}^{\alpha\beta}}{\partial (u_i^\gamma - u_j^\gamma)} \right\rvert_{\ve{u}_i - \ve{u}_j = 0} (u_i^\gamma - u_j^\gamma).
\label{eq:Aijexpansion}
\end{equation}

\noindent If we insert the expansion (\ref{eq:Aijexpansion}) into the Hamiltonian \eqref{eq:H3origin}, we see that a new term coupling the spins at site $i$ and $j$ to the relative displacement appears. This interaction is proportional to the coupling tensor $\partial A_{ij}^{\alpha\beta} / \partial (u_i^\gamma - u_j^\gamma) \rvert_{\ve{u}_i - \ve{u}_j = 0}$. Now note that this coupling tensor is not a tensor of constant coefficients, like $A_{ij}^{\alpha\beta}$. It transforms differently due to the operator $\partial /\partial (u_i^\gamma - u_j^\gamma)$ working on it. For instance, under inversion, ${\partial /\partial (u_i^\gamma - u_j^\gamma) \rightarrow - \partial/\partial (u_i^\gamma - u_j^\gamma)}$. For convenience, we therefore introduce the coupling tensor $B_{ijkl}^{\alpha\beta\gamma\lambda}$, implicitly defined by

\begin{equation}
\left. \frac{\partial A_{ij}^{\alpha\beta}}{\partial (u_i^\gamma - u_j^\gamma)} \right\rvert_{\ve{u}_i - \ve{u}_j = 0} \equiv \sum_{kl} \sum_{\lambda} B_{ijkl}^{\alpha\beta\gamma\lambda} (r_k^\lambda - r_l^\lambda),
\end{equation}

\noindent where $k$ and $l$ run over all lattice sites. The components of $B_{ijkl}^{\alpha\beta\gamma\lambda}$ transform just as scalars under inversion, and we therefore choose to proceed with $B_{ijkl}^{\alpha\beta\gamma\lambda}$ as the coupling tensor. As before, we note that $(r_k^\gamma - r_l^\gamma)$ is not a dynamical variable, but merely a constant with the desired transformation properties once the lattice has been defined. Its introduction does therefore not alter the physical content of the Hamiltonian.

We find the non-zero tensor elements $B_{ijkl}^{\alpha\beta\gamma\lambda}$ by requiring the Hamiltonian \eqref{eq:H3origin} to respect the symmetries of the lattice. Following the argument presented in Appendix \ref{sec:app1}, we may pick only the $B_{ijkl}^{\alpha\beta\gamma\lambda}$'s where $k,l \in \{ i,j \}$ to be non-zero. We define $B_{ij}^{\alpha\beta\gamma\lambda} \equiv B_{ijii}^{\alpha\beta\gamma\lambda}$, and may then write the exchange Hamiltonian to first order in the relative displacement as

\begin{equation}
\mathcal{H}_{s} = \sum_{ij} \sum_{\alpha\beta} A_{ij}^{\alpha\beta}(0) S_i^\alpha S_j^\beta + \sum_{ij} \sum_{\alpha\beta\gamma\lambda} B_{ij}^{\alpha\beta\gamma\lambda}  S_i^\alpha S_j^\beta (u_i^\gamma - u_j^\gamma) (r_i^\lambda - r_j^\lambda).
\end{equation}

\noindent We recognize that the second term is equivalent to the Hamiltonian \eqref{eq:Hsimple3_2} after symmetrization in the $\gamma\lambda$ indices in order to disregard rotations. 

\section{Long-wavelength magnetoelastic Hamiltonian}
\label{sec:app3}

We will now derive the long-wavelength magnetoelastic Hamiltonian for a cubic collinear antiferromagnets, starting from the general form given in Eq. \eqref{eq:HME2}. In a collinear antiferromagnet, there are two sublattices, A and B. In the (classical) ground state, the spins on each sublattice are anti-parallel. Motivated by the bipartite lattice, we first separate the sums over $i$ and $\ve{\delta}$ in Eq. \eqref{eq:HME2} into two separate contributions, one in which $i$ is a site on sublattice A, and one in which $i$ is a site on sublattice B. We can further separate each of these sums into two sums, where the vector $\ve{\delta}$ points between sites on the same sublattice, or between the sublattices. In total, we thus have four separate sums. However, because sublattice A and B are equivalent, there are only two independent sums. We therefore have

\begin{equation}
\mathcal{H}_{\rm ME} = 2 \sum_{\alpha\beta} \sum_{i \in A} S_{i}^{\alpha} \left( \sum_{\ve{\delta} \in \mathrm{AA}} B_{\abs{\ve{\delta}}}^{\alpha\beta} S_{i + \ve{\delta}}^{\beta} R_{i,i+\ve{\delta}}^{\alpha\beta} +
\sum_{\ve{\delta} \in \mathrm{AB}} B_{\abs{\ve{\delta}}}^{\alpha\beta} S_{i + \ve{\delta}}^{\beta} R_{i,i+\ve{\delta}}^{\alpha\beta} \right) \, .
\label{eq:afm1}
\end{equation}

\noindent where $\ve{\delta} \in$ AA (AB) denotes that $\ve{\delta}$ points from a site on sublattice A to a site on sublattice A (B). We will now transit to the long-wavelength limit. First define two spin fields, $\ve{S}^{\rm A}(\ve{r})$ and $\ve{S}^{\rm B}(\ve{r})$, living on sublattice A and B respectively. We then introduce the Néel vector $\ve{n}(\ve{r}) = \frac{1}{2} \left( \ve{S}^{\rm A}(\ve{r}) - \ve{S}^{\rm B}(\ve{r}) \right)$ and the local magnetization $\ve{m}(\ve{r}) = \frac{1}{2} \left( \ve{S}^{\rm A}(\ve{r}) + \ve{S}^{\rm B}(\ve{r}) \right)$. We will in the following neglect the local magnetization, and may therefore express the Hamiltonian as

\begin{align}
\begin{split}
\mathcal{H}_{\rm ME} = \sum_{\alpha\beta} \sum_{i \in A} n^{\alpha}(\ve{r_i}) &\Bigg( \sum_{\ve{\delta} \in \mathrm{AA}} B_{\abs{\ve{\delta}}}^{\alpha\beta} n^{\beta}(\ve{r_i} + \ve{\delta}) R_{i,i+\ve{\delta}}^{\alpha\beta} \\
&\quad -
\sum_{\ve{\delta} \in \mathrm{AB}} B_{\abs{\ve{\delta}}}^{\alpha\beta} n^{\beta}(\ve{r_i} + \ve{\delta}) R_{i,i+\ve{\delta}}^{\alpha\beta} \Bigg) \, .
\label{eq:afm1}
\end{split}
\end{align}

\noindent Notice the relative minus sign between the sum over AA and AB, which has been introduced because the spins on the two sublattices are approximately anti-parallel, $\ve{n} = \ve{S}^{\rm A}(\ve{r}) \approx -\ve{S}^{\rm B}(\ve{r})$. This sign change is characteristic to antiferromagnets and strongly affects the final result in certain materials.

In order to fully transit to the long-wavelength limit, we first need to do the sum over $\ve{\delta}$. In order to do so, we expand the Néel vector to first order, 

\begin{equation}
n^\alpha(\ve{r}_i + \ve{\delta}) \approx n^\alpha(\ve{r}_i) + \sum_\gamma \left. \frac{\partial n^\alpha(\ve{r})}{\partial r^\gamma} \right\rvert_{\ve{r} = \ve{r}_i} \delta^\gamma.
\end{equation}

\noindent For notational simplicity, we define $\left. \frac{\partial n^\alpha(\ve{r})}{\partial r^\gamma} \right\rvert_{\ve{r} = \ve{r}_i} \equiv D^{\alpha\gamma}(\ve{r_i})$. The Hamiltonians then reads

\begin{align}
\begin{split}
\mathcal{H}_{\rm ME} &= \sum_{\alpha\beta} \sum_{i \in A} n^{\alpha}(\ve{r_i}) n^{\beta}(\ve{r_i}) \Bigg(  \sum_{\ve{\delta} \in \mathrm{AA}} B_{\abs{\ve{\delta}}}^{\alpha\beta}  R_{i,i+\ve{\delta}}^{\alpha\beta} - \sum_{\ve{\delta} \in \mathrm{AB}} B_{\abs{\ve{\delta}}}^{\alpha\beta} R_{i,i+\ve{\delta}}^{\alpha\beta}  \Bigg) \\ 
&\quad + \sum_{\alpha\beta} \sum_{i \in A} n^{\alpha}(\ve{r_i}) n^{\beta}(\ve{r_i}) D^{\alpha\gamma}(\ve{r}_i) \\
&\quad \times \Bigg(  \sum_{\ve{\delta} \in \mathrm{AA}} B_{\abs{\ve{\delta}}}^{\alpha\beta}  R_{i,i+\ve{\delta}}^{\alpha\beta} \delta^\gamma - \sum_{\ve{\delta} \in \mathrm{AB}} B_{\abs{\ve{\delta}}}^{\alpha\beta} R_{i,i+\ve{\delta}}^{\alpha\beta} \delta^\gamma  \Bigg).
\label{eq:afm1}
\end{split}
\end{align}

\noindent Now rearrange the sums over $\ve{\delta}$ into separate sums arising from the different layers of $n$'th nearest neighbors at distances $|\ve{\delta}_n|$. That is, write

\begin{align}
\sum_{\ve{\delta}} B_{\abs{\ve{\delta}}}^{\alpha\beta}  R_{i,i+\ve{\delta}}^{\alpha\beta} &= B_{\abs{\ve{\delta}_1}}^{\alpha\beta} \sum_{\ve{\delta}_1} R_{i,i+\ve{\delta}_1}^{\alpha\beta} + B_{\abs{\ve{\delta}_2}}^{\alpha\beta} \sum_{\ve{\delta}_2} R_{i,i+\ve{\delta}_2}^{\alpha\beta} + ... \; , \\
\sum_{\ve{\delta}} B_{\abs{\ve{\delta}}}^{\alpha\beta}  R_{i,i+\ve{\delta}}^{\alpha\beta} \delta^\gamma &= B_{\abs{\ve{\delta}_1}}^{\alpha\beta} \sum_{\ve{\delta}_1} R_{i,i+\ve{\delta}_1}^{\alpha\beta} \delta_1^\gamma + B_{\abs{\ve{\delta}_2}}^{\alpha\beta} \sum_{\ve{\delta}_2} R_{i,i+\ve{\delta}_2}^{\alpha\beta} \delta_2^\gamma + ... \;.
\end{align}

\noindent Note that we have put $B_{\abs{\ve{\delta}}}^{\alpha\beta}$ outside the sums, as these coefficients are equal for all neighbors included in an $n$'th nearest neighbor summation. We now need to evaluate only two kinds of sums, namely $\sum_{\ve{\delta}_n} R_{i,i+\ve{\delta}_n}^{\alpha\beta}$ and $\sum_{\ve{\delta}_n} R_{i,i+\ve{\delta}_n}^{\alpha\beta} \delta_n^\gamma$.

In cubic collinear antiferromagnets, there are three types of $n$'th nearest neighbor layers to any given spin $i$ on sublattice A: 1. All neighbors belong to sublattice A and are therefore parallel to $\ve{S}_i$. 2. All neighbors belong to sublattice B, and are therefore anti-parallel to $\ve{S}_i$. 3. Half of the spins belong to sublattice A and the other half belong to sublattice B. In situation 1 and 2, all $n$'th nearest neighbors belong to one sublattice. We will now assume that only the $B_{\abs{\ve{\delta}_n}}^{\alpha\beta}$'s where $|\ve{\delta}_n|$ is much shorter than the wavelength of lattice strains contributes to the sum. In other words, we assume that the range of the magnetoelastic interaction is relatively short. In cubic crystals, we then have to good approximation $\sum_{\ve{\delta}_n}  R_{i,i+\ve{\delta}_n}^{\alpha\beta} = C_n \epsilon^{\alpha\beta}(\ve{r}_i)$ where $C_n$ is some constant, and $\sum_{\ve{\delta}_n}  R_{i,i+\ve{\delta}_n}^{\alpha\beta} \ve{\delta}^\gamma_n = 0$. Situation 1 and 2 thus couple elements of the Néel field straightforwardly to elements of the conventional strain tensor.

In situation 3, however, half of the terms in the summation over $\ve{\delta}_n$ comes with a minus sign, as is evident from Eq. \eqref{eq:afm1}. By evaluating the sums over $\ve{\delta}$ arising from situation 3 explicitly, we find $\sum_{\ve{\delta}_n}'  R_{i,i+\ve{\delta}_n}^{\alpha\beta} = \tilde{C}_n \tilde{\epsilon}^{\alpha\beta}(\ve{r}_i)$ where the primed summation indicates that terms arising from a $\ve{\delta}_n$ pointing between the sublattices are accompanied with a minus sign, $\tilde{C}_n$ is some constant, and $\tilde{\epsilon}^{\alpha\beta}(\ve{r}_i)$ is some tensor with elements that are linear combinations of strain tensor elements. In general, the structure of this tensor depends on the microscopic spin structure of the material. The last sum gives no contribution for situation 3 either, $\sum_{\ve{\delta}_n}'  R_{i,i+\ve{\delta}_n}^{\alpha\beta} \ve{\delta}^\gamma_n = 0$, meaning that the gradient of the Néel field does not contribute to the final Hamiltonian to this order. Situation 3 hence couples the Néel field to another tensor $\tilde{\epsilon}$, not equal to the strain tensor. This tensor only arises when there exist nearest neighbor layers in which a portion of the spins are part of sublattice A and a portion are part of sublattice B. NiO is an example of such a material, and likewise are CoO, FeO and MnO.

We may now finalize the transition into the long-wavelength limit, by taking ${\sum_i \rightarrow \int \mathrm{d}\ve{r}}$. The magnetoelastic Hamiltonian for cubic collinear antiferromagnets then follows as

\begin{equation}
\mathcal{H}_{\rm ME}^{\rm AFM} = \sum_{\alpha\beta} \int \mathrm{d}\ve{r} \; n^\alpha(\ve{r}) n^\beta(\ve{r}) \left[ B^{\alpha\beta}  \epsilon^{\alpha\beta}(\ve{r}) + \tilde{B}^{\alpha\beta} \tilde{\epsilon}^{\alpha\beta}(\ve{r}) \right].
\label{eq:me_afm}
\end{equation}

\noindent where $B^{\alpha\beta} = \sum_{\mathrm{s}(n) \in \{1,2\}} C_n B_{|\ve{\delta}_n|}^{\alpha\beta}$ and $\tilde{B}^{\alpha\beta} = \sum_{\mathrm{s}(n) \in 3} \tilde{C}_n B_{|\ve{\delta}_n|}^{\alpha\beta}$, where the notation $\mathrm{s}(n) \in \{1,2\}$ means summing over the $n$'th nearest neighbor layers belonging to either situation 1 or 2 as introduced in the above paragraph, and $\mathrm{s}(n) \in 3$ means summing over the $n$'th nearest neighbor layers belonging to situation 3. Each tensor contains two independent coefficients, $B^{\alpha\beta} = B^{\parallel} \delta^{\alpha\beta} + B^{\perp} \left( 1 - \delta^{\alpha\beta} \right)$ and $\tilde{B}^{\alpha\beta} = \tilde{B}^{\parallel} \delta^{\alpha\beta} + \tilde{B}^{\perp} \left( 1 - \delta^{\alpha\beta} \right)$.


%

\end{document}